# Large thermo-spin effects in Heusler alloy based spin-gapless semiconductor thin films

Amit Chanda[1*], Deepika Rani[2], Derick DeTellem[1], Noha Alzahrani[1], Dario A. Arena[1], Sarath Witanachchi[1], Ratnamala Chatterjee[2], Manh-Huong Phan[1] and Hariharan Srikanth[1*]

[1] Department of Physics, University of South Florida, Tampa FL 33620

[2] Physics Department, Indian Institute of Technology Delhi, New Delhi - 110016

*Corresponding authors: achanda@usf.edu; sharihar@usf.edu

## Abstract

Recently, Heusler alloys-based spin gapless semiconductors (SGSs) with high Curie temperature ($T_C$) and sizeable spin polarization have emerged as potential candidates for tunable spintronic applications. We report comprehensive investigation of the temperature dependent ANE and intrinsic longitudinal spin Seebeck effect (LSSE) in CoFeCrGa thin films grown on MgO substrates. Our findings show the anomalous Nernst coefficient for the MgO/CoFeCrGa (95 nm) film is $\approx 1.86\ \mu\text{V}\cdot\text{K}^{-1}$ at room temperature which is nearly two orders of magnitude higher than that of the bulk polycrystalline sample of CoFeCrGa ($\approx 0.018\ \mu\text{V}\cdot\text{K}^{-1}$) but comparable to that of the magnetic Weyl semimetal $Co_2MnGa$ thin film ($\approx 2-3\ \mu\text{V}\cdot\text{K}^{-1}$). Furthermore, the LSSE coefficient for our MgO/CoFeCrGa(95nm)/Pt(5nm) heterostructure is $\approx 20.5\ \text{nV}\cdot\text{K}^{-1}\cdot\Omega^{-1}$ at room temperature which is twice larger than that of the half-metallic ferromagnetic $La_{0.7}Sr_{0.3}MnO_3$ thin films ($\approx 9\ \text{nV}\cdot\text{K}^{-1}\cdot\Omega^{-1}$). We show that both ANE and LSSE coefficients follow identical temperature dependences and exhibit a maximum at $\approx 225$ K which is understood as the combined effects of inelastic magnon scatterings and reduced magnon population at low temperatures. Our analyses not only indicate that the extrinsic skew scattering is the dominating mechanism for ANE in these films but also provide critical insights into the functional form of the observed temperature dependent LSSE at low temperatures. Furthermore, by employing radio frequency transverse susceptibility and broadband ferromagnetic resonance in combination with the LSSE measurements, we establish a correlation among the observed LSSE signal, magnetic anisotropy and Gilbert damping of the CoFeCrGa thin films, which will be beneficial for fabricating tunable and highly efficient Heusler alloys based spincaloritronic nanodevices.

# 1. INTRODUCTION

The past few years have witnessed extensive research efforts in the field of spincaloritronics for the development of highly efficient next-generation spin-based electronic devices by combining the versatile advantages of spintronics and thermoelectricity, with the aim of finding novel avenues for waste heat recovery and thermoelectric energy conversion[1,2]. Fundamental knowledge of the interplay between heat, charge, and spin degrees of freedom not only allowed us to understand how thermal gradients can be utilized to manipulate and control the flow of spin angular momenta inside a material at nanoscale, but also helped the scientific community to explore various intriguing thermo-spin transport phenomena, such as the anomalous Nernst effect (ANE)[3], spin Nernst effect[4], spin Seebeck effect[5,6], spin Peltier effect[7] and so on.

The ANE refers to the generation of a transverse thermoelectric voltage in a magnetic conductor/semiconductor by the application of a thermal gradient and an external magnetic field[8,9]. The ANE has been observed in a large range of magnetic materials, from half-metallic ferromagnets such as hole-doped manganites[10], cobaltites[11–13], spin gapless semiconductors[14] to ferrimagnets such as iron oxide[15], Mn-based nitride[16], as well as unconventional magnetic systems with topologically non-trivial phases such as topological full Heusler ferromagnets[3,17–19], ferromagnetic Weyl semimetals[20,21], two-dimensional topological van der Waals ferromagnets[22,23], chiral[8] and canted[24] topological antiferromagnets *etc*. In a topological magnetic material, charge carriers moving through a periodic potential with strong spin-orbit coupling (SOC) acquire an additional anomalous velocity perpendicular to their original trajectory due to the non-zero Berry curvature at the Fermi level[25]. This anomalous velocity causes a real space spin selective deflection of the charge carriers and leads to a potentially large ANE response in these topological magnetic materials compared to conventional magnets[25]. In addition to the aforementioned intrinsic origin, ANE can also originate from extrinsic effects for example, asymmetric skew scattering of charge carriers as observed in Heusler ferromagnets[14,26,27], hole-doped manganites[10], cobaltites[11–13], spin gapless semiconductors[14], iron oxide[15] etc.

On the other hand, the longitudinal spin Seebeck effect (LSSE) refers to the thermal generation of magnonic spin current in a ferromagnetic (FM) material by the concurrent applications of a temperature gradient and an external magnetic field across a FM/heavy metal (HM) bilayer structure and injection of that spin current to the adjacent HM layer with strong SOC, which is then converted into electrically detectable charge current in the HM layer via the inverse spin Hall effect (ISHE)[1,28–30]. The bilayer structure consisting of the ferrimagnetic insulator $Y_3Fe_5O_{12}$ (YIG) and Pt is known as the benchmark system for generating pure spin current and hence, LSSE[28,30–34]. Apart from YIG, other magnetic insulators for example, the compensated ferrimagnetic insulator $Gd_3Fe_5O_{12}$,[35,36] insulating spinel ferrites $CoFe_2O_4$, $NiFe_2O_4$[37,38], noncollinear antiferromagnetic insulator $LuFeO_3$[39] etc., have also emerged as promising spincaloritronic materials. Nevertheless, observation of LSSE is not only restricted to magnetic insulators, but it has also been observed in metallic[5,40], half-metallic[41–43] and semiconducting ferromagnets[44].

Although ANE and LSSE are two distinct types of magnetothermoelectric phenomena, they share common origin for materials exhibiting extrinsic effects dominated ANE[45]. In both the cases, simultaneous application of thermal gradient and external magnetic field generates magnonic excitations. While in the case of ANE, the thermally generated magnons transfer spin angular momenta to the itinerant electrons of the FM via the electron-magnon scattering and thereby dynamically spin polarizes them, in the case of LSSE, a spatial gradient of those thermally generated magnons leads to magnon accumulation close to the FM/HM interface and pumps spin current to the HM layer[45]. Large magnon-induced ANE has been observed in MnBi single crystal[45]. However, observation of large ANE in a FM conductor does not necessarily indicate a promise for a large LSSE, and vice versa. Therefore, it would be technologically advantageous from the perspective of spincaloritronic device applications and thermal energy harnessing to search for a FM material that can simultaneously exhibit large LSSE and ANE.

In recent years, Heusler alloys-based spin gapless semiconductors (SGSs) have emerged as promising magnetic materials for tunable spintronic applications as they not only combine the characteristics of both half-metallic ferromagnets and gapless semiconductors,[46] but also possess high Curie temperature ($T_C$) and substantial spin polarization[47–50]. We have recently observed large ANE in the bulk sample of Heusler alloy based SGS: CoFeCrGa with $T_C \approx 690$ K, [14,50,51] which was the first experimental observation of ANE in the SGS family. Our fascinating observation motivated us to explore ANE as well as LSSE in the CoFeCrGa thin films. Although SGS has been theoretically predicted to be a promising candidate for spintronic applications[52], there is no previous experimental study on the thermo-spin transport phenomena, especially LSSE in SGS thin films. In this paper, we report on the temperature dependent ANE and LSSE in the CoFeCrGa single layer and CoFeCrGa/Pt bilayer films with different CoFeCrGa film thicknesses. We found that both ANE and LSSE coefficients follow identical temperature dependences and exhibit a maximum at $\approx 225$ K which is understood as the combined effects of inelastic magnon scatterings and reduced magnon population at low temperatures. Our analyses not only indicate that the extrinsic skew scattering is the dominating mechanism for ANE in these films but also provide critical insights into the functional form of the observed temperature dependent LSSE. Furthermore, we have established a correlation among the observed LSSE signal, magnetic anisotropy and Gilbert damping of the CoFeCrGa thin films which will be beneficial for fabricating tunable and efficient spincaloritronic devices.

## 2. EXPERIMENTAL SECTION

The thin films of CoFeCrGa were grown on single crystal MgO (001) substrates of surface area $5 \times 5$ mm$^2$ using an excimer KrF pulsed laser deposition (PLD) system. The films were deposited at 500 °C and were further annealed in-situ at 500 °C for 30 min to further enhance the chemical order and crystallization. The film surface morphology was investigated by field emission gun – scanning electron microscopy (FEG-SEM) and atomic force microscopy (AFM), while the structural properties of the thin films were identified by x-ray diffraction (XRD) using monochromatic Cu Kα radiation.

AFM and temperature dependent magnetic force microscopy (MFM) measurements were performed on a Hitachi 5300E system. All measurements were done under high vacuum (P ≤ 10$^{-6}$ Torr). MFM measurements utilized PPP-MFMR tips, which were magnetized out-of-plane with respect to the tip surface via a permanent magnet. Films were first magnetized to their saturation magnetization by being placed in a 1T static magnetic field, in-plane with the film surface. After that AC demagnetization of the film was implemented before initiating the MFM scans. After scans were performed, first a linear background was subtracted which comes from the film not being completely flat on the sample stage. After that, a parabolic background was subtracted, which arises from the nonlinear motion of the piezoelectric crystal that drives the *x-y* translation. Phase standard deviation was determined by fitting a Gaussian to the image phase distribution and extracting the standard deviation from the fit parameters.

The DC magnetic measurements on the samples at temperatures between 100 K and 300 K were performed using a vibrating sample magnetometer (VSM) attached to a physical property measurement system (PPMS), Quantum Design. A linear background originating from the diamagnetic MgO substrate was thereby subtracted. Due to a trapped remanent field inside the superconducting coils, the measured magnetic field was corrected using a paramagnetic reference sample.

The longitudinal electrical resistivity, longitudinal Seebeck coefficient, and thermal conductivity of the bulk samples were simultaneously measured with the thermal transport option (TTO) of the PPMS. The electrical resistivity and Hall measurements on the thin film samples were performed using the DC resistivity option of the PPMS by employing a standard four point measurement technique with sourcing currents of 500 μA and 1 mA, respectively.

The temperature dependence of the effective magnetic anisotropy fields of the MgO/CoFeCrGa films were measured by using a radio frequency (RF) transverse susceptibility (TS) measurement technique that exploits a self-resonant tunnel diode oscillator (TDO) circuit with a resonance frequency of ≈12 MHz[53,54]. The PPMS was used as a platform to sweep the external DC magnetic field and temperature. During the TS measurement, the MgO/CoFeCrGa thin film samples were firmly placed inside an inductor coil (L), which is a component of an LC resonator circuit. The coil containing the sample was positioned at the base of the PPMS sample chamber through a multifunctional PPMS probe in such a way that the axial RF magnetic field generated inside the coil stayed parallel to the film surface, but perpendicular to the DC magnetic field generated by the superconducting magnet of the PPMS. In presence of both the RF and DC magnetic fields, the dynamic transverse susceptibility of the sample changes which eventually changes the resonance frequency of the LC circuit[53]. From the magnetic field dependence of the shift in the resonance frequency recorded by an Agilent frequency counter, we obtained the field dependent transverse susceptibility.

The ANE and LSSE measurements were performed using a custom-designed setup assembled on a universal PPMS sample puck, as shown in our previous reports[14,36]. For both the measurements, the thin film samples were sandwiched between two copper plates. A single layer of thin Kapton tape was thermally anchored to the bare surfaces of the top (cold) and bottom (hot) copper plates. Cryogenic Apiezon N-grease was used to create good thermal connectivity between the thin film surface and that of the Kapton tapes. A resistive heater (PT-100 RTD sensor) and a calibrated Si-diode thermometer (DT-621-HR silicon diode sensor) were attached to each of those copper plates. The temperatures of both these copper plates were monitored and controlled individually by employing two distinct separate temperature controllers (Scientific Instruments Model no. 9700). The top copper plate was thermally linked to the base of the PPMS universal puck using a pair of molybdenum screws and a 4 mm thick Teflon block was thermally sandwiched between the universal PPMS puck base and the bottom copper plate to maintain a temperature difference of ~ 10 K between the hot copper plate and the PPMS universal puck base. The Ohmic contacts for the ANE and LSSE voltage measurements were made by using a pair of thin gold wires of 25 µm diameter to the Pt layer by high quality conducting silver paint (SPI Supplies). In presence of an applied temperature gradient along the $z$-direction, and an in-plane external DC magnetic field applied along the $x$-direction, the transverse thermoelectric voltage generated along the $y$-direction across the Pt layer due to the ISHE ($V_{ISHE}$) and across the CoFeCrGa film itself due to the ANE was recorded with a Keithley 2182a nanovoltmeter.

Broadband ferromagnetic resonance (FMR) measurements were performed using a broadband FMR spectrometer (NanOsc^TM Phase-FMR, Quantum Design Inc., USA) integrated to the Dynacool PPMS[55].

## 3. RESULTS AND DISCUSSION

### 3.1. Structural and morphological properties

**Figure 1**(a) shows the X-ray 2θ-ω (out of plane) diffraction pattern for CoFeCrGa (95nm) film grown on MgO (001) substrate. In addition to the peaks corresponding to the MgO substrate, there are additional (002) and (004) diffraction peaks from the film, indicating the growth in the (001) orientation. The formation of B2 CoFeCrGa structure is confirmed by the presence of (002) peak. To find the CoFeCrGa (220) peak intensity, a 2θ-θ scan was performed with χ= 45° as shown in **Fig. 1**(b). The lattice parameter as estimated by applying the Bragg equation to the (022) peak, was found to be 5.76 Å.

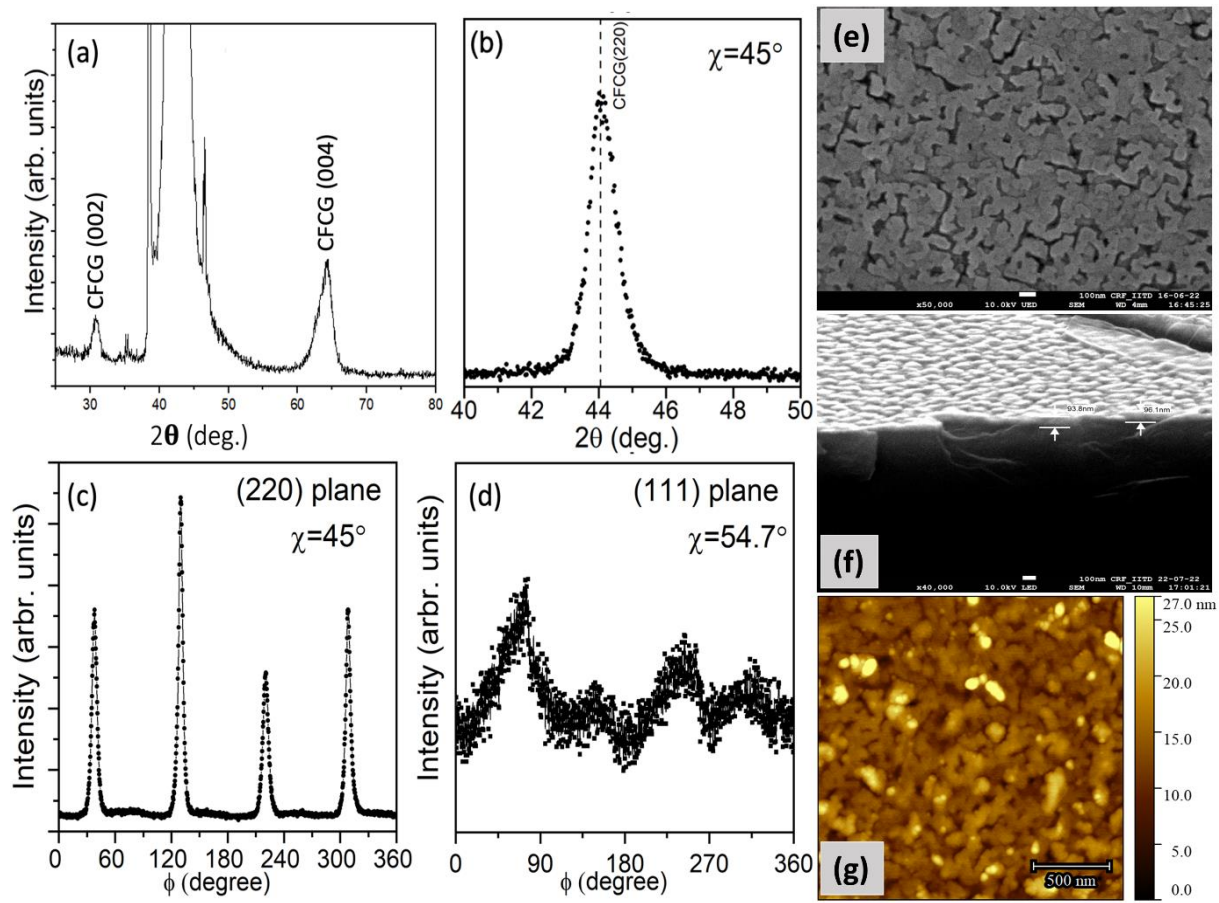

**Figure 1.** (a) XRD of MgO/CoFeCrGa(95nm) film: ω–2θ (out-of-plane) scan. (b) The 2θ–θ scan of the (022) plane. (c) Phi-scan of the (022) plane. (d) $\phi$-scan of the (111) plane. (e) FEG-SEM, (f) Cross sectional SEM image and (g) AFM image for the MgO/CoFeCrGa(95nm) film.

To further confirm the epitaxial growth of the CoFeCrGa (95nm) film, $\phi$-scan was performed for the (220) and (111) planes by tilting the sample, i.e., $\chi= 45°$ for the (220) plane and $\chi= 54.7°$ for the (111) plane (**Fig. 1**(c), and **Fig. 1**(d)). The $\phi$ - scans of both (220) and (111) plane show a four-fold symmetry, as four well defined peaks periodically separated from each other by 90° were observed. The presence of both (111) and (200) peaks rule out the possibility of complete A2 or B2 disorder, however partial disorder can still be present. The chemical composition was interpreted by the scanning electron microscopy energy-dispersive spectroscopy (SEM-EDS) measurements and was found to be $Co_{1.05}Fe_{1.05}Cr_{0.9}Ga_{0.99}$, which is very close to the ideal stoichiometric composition expected for an equiatomic quaternary Heusler alloy. The surface morphology of the film obtained from the FEG-SEM image is shown in **Fig. 1**(e), which indicates that the film is homogenous, which was further confirmed by AFM measurements as shown in **Fig. 1**(g). A low root-mean-square (RMS) roughness of ≈ 2.5 nm is achieved for the CoFeCrGa film, as noticeable in the AFM image shown in **Fig. 1**(g). The cross-section SEM image of the film is shown in **Fig. 1**(f), which indicates that the film thickness (~95± 5 nm).

In **Fig. 2**, we show the temperature dependent magnetic force microscopy (MFM) images recorded on the MgO/CoFeCrGa(95nm) film. The MFM image at 300 K (see **Fig. 2**(a)) shows a bright/dark contrast with highly irregular shaped features indicating cloudlike domain-clusters[56]. Note that in MFM, the domain-image contrast is determined by the magnetic force-gradient $\left(\frac{dF}{dz}\right)$ between the sample and the MFM tip (magnetized ⊥ to the film-surface), which is proportional to the perpendicular component of the stray field of the film[57,58]. For our film, due to low bright/dark contrast patterns of the MFM images in the *T*-range: 160K $\leq T \leq$ 300 K (**Fig. 2**(a)-(e)), the domain boundaries are not as well-defined as observed in films with strong PMA[59].

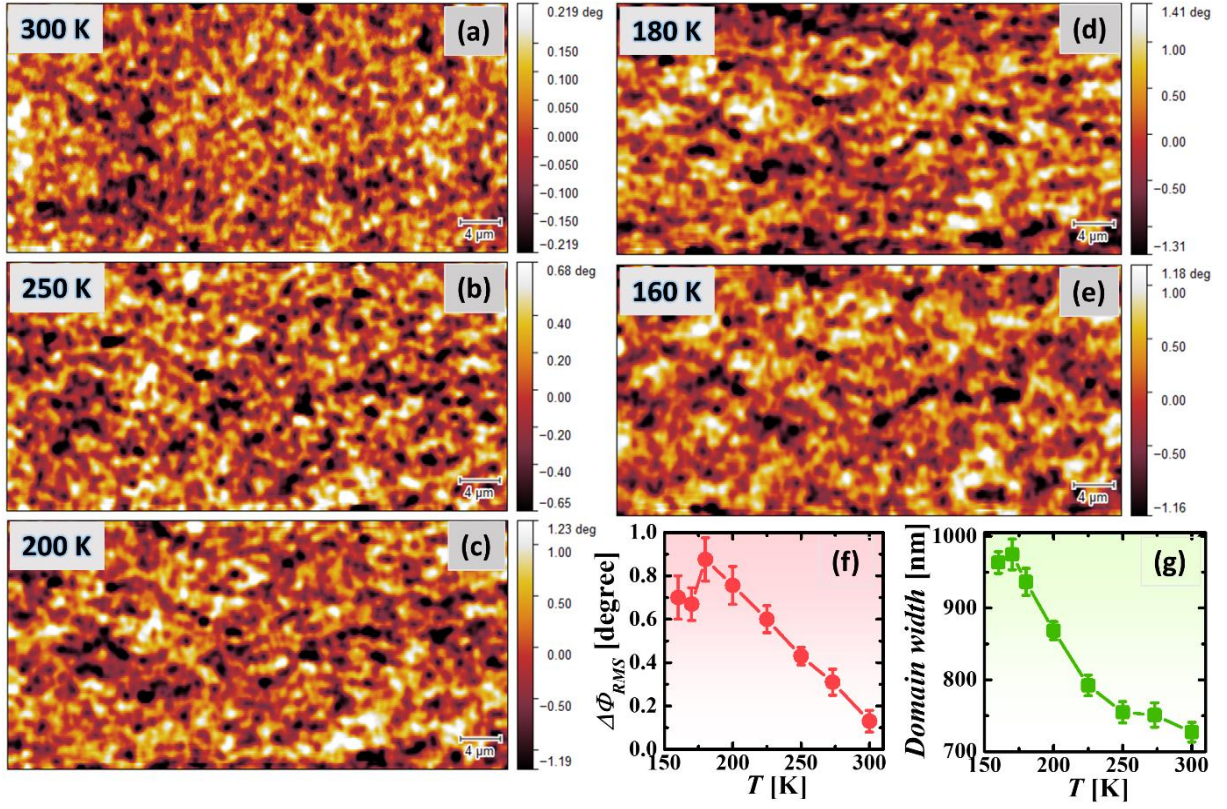

**Figure 2.** Magnetic force microscopy (MFM) images of MgO/CoFeCrGa(95nm) film measured at **(a)** $T = 300$ K, **(b)** $T = 250$ K, **(c)** $T = 200$ K, **(d)** $T = 180$ K, and **(e)** $T = 160$ K while cooling the sample after applying an IP magnetic field (higher than the IP saturation field) and then AC demagnetization of the sample at 300 K. **(f)** The RMS value of the phase shift, $\Delta\phi_{RMS}$ as a function of temperature for the MgO/CoFeCrGa(95nm) film extracted from the MFM images. (g) The average domain width as a function of temperature obtained from the MFM images.

A steep increase in the root mean square (RMS) value of the phase shift,[57] $\Delta\phi_{RMS} \approx \frac{Q}{K}\left[\frac{dF}{dz}\right]$ (Q = quality factor and $K$ = spring constant of the tip; hence, $\Delta\phi_{RMS} \propto$ average domain contrast[58]) has also been observed below 300 K (see **Fig. 2**(f)). However, $\Delta\phi_{RMS}$ decreases slightly below 180 K. Average domain widths were determined by calculating the 2D autocorrelation across the MFM images, then determining the full-width half-max (FWHM) of arbitrary lines through the 2D autocorrelation spectra. As shown in **Fig.** 2(g), the average domain width also increases with decreasing temperature followed by a slight decrease below 170 K.

## 3.2. Magnetic and electrical transport properties

Previous studies on bulk CoFeCrGa[50,51] as well as MgO/CoFeCrGa thin films[60] reveal that the ferromagnetic transition temperature of this sample is very high (at least $\geq$ 500 K). The main panel of **Fig. 3**(a) shows the magnetic field dependence of magnetization, $M(H)$ of our MgO/CoFeCrGa film measured at selected temperatures in the range: 125 K $\leq T \leq$ 300 K in presence of an in-plane sweeping magnetic field. The $M(H)$ loops exhibit very small coercivity throughout the measured temperature range.

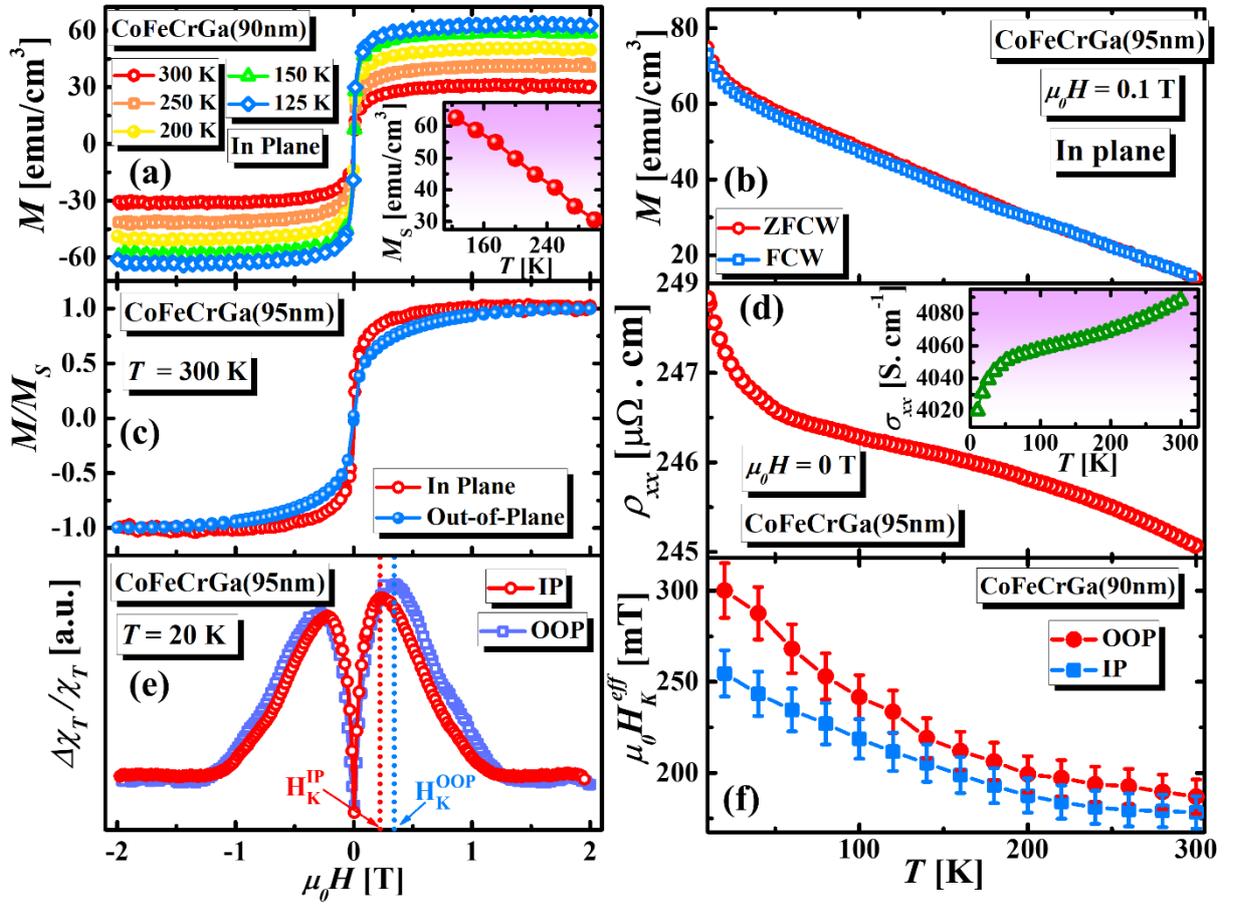

**Figure 3.** (a) Main panel: magnetic field dependence of magnetization, $M(H)$ of our MgO/CoFeCrGa(95nm) film measured at selected temperatures in the range: 125 K $\leq T \leq$ 300 K in presence of an in-plane sweeping magnetic field, inset: temperature dependence of the saturation magnetization, $M_S$. (b) Temperature dependence of magnetization, $M(T)$ measured in zero-field-cooled warming (ZFCW) and field-cooled-warming (FCW) protocols in presence of an external magnetic field: $\mu_0 H = 0.1$ T. (c) Normalized $M(H)$ hysteresis loops at $T = 300$ K for the in-plane (IP) and out-of-plane (OOP) configurations. (d) Main panel:

temperature dependence of longitudinal resistivity, $\rho_{xx}(T)$ for the MgO/CoFeCrGa film in the temperature range: $10\,\text{K} \leq T \leq 300\,\text{K}$, inset shows corresponding temperature dependence of electrical conductivity, $\sigma_{xx}(T)$. (e) The bipolar field scans ($+H_{DC}^{max} \to -H_{DC}^{max} \to +H_{DC}^{max}$) of $\frac{\Delta \chi_T}{\chi_T}(H_{DC})$ for MgO/CoFeCrGa(95nm) film measured at $T = 20\,\text{K}$ for both IP ($H_{DC}$ is parallel to the film surface) and OOP ($H_{DC}$ is perpendicular to the film surface) configurations. (f) Temperature variations of the effective anisotropy fields: $H_K^{IP}$ and $H_K^{OOP}$ for our MgO/CoFeCrGa(95nm) film.

As shown in the inset of **Fig. 3**(a), the saturation magnetization, $M_S$ increases almost linearly with decreasing temperature, which is in agreement with the temperature dependent $\Delta\phi_{RMS}$ obtained from the MFM images[58]. In **Fig. 3**(b), we show the temperature dependence of magnetization, $M(T)$ measured in zero-field-cooled warming (ZFCW) and field-cooled-warming (FCW) protocols in presence of an external magnetic field: $\mu_0 H = 0.1\,\text{T}$. It is evident that both ZFCW and FCW $M(T)$ increases with decreasing temperature down to 10 K below which it shows a slight up-turn. Furthermore, the ZFCW and FCW $M(T)$ curves do not exhibit any considerable bifurcation at low temperatures which is indicative of the absence of any glassy magnetic ground state. **Fig. 3**(c) shows the normalized $M(H)$ hysteresis loops at $T = 300\,\text{K}$ for the in-plane (IP) and out-of-plane (OOP) configurations confirming the soft ferromagnetic nature of the film along the IP direction, which is consistent with a recent report on this system[60].

The main panel of **Fig. 3**(d) demonstrates the $T$-dependence of longitudinal resistivity, $\rho_{xx}(T)$ for the MgO/CoFeCrGa film in the temperature range: $10\,\text{K} \leq T \leq 300\,\text{K}$. It is obvious that $\rho_{xx}(T)$ exhibits semiconducting-like resistivity $\left(\frac{\partial \rho_{xx}}{\partial T} > 0\right)$ throughout the temperature range. The inset of **Fig. 3**(d) shows the $T$-dependence of electrical conductivity, $\sigma_{xx}(T)$ for the MgO/CoFeCrGa film. Note that the values of both $\rho_{xx}(T)$ and $\sigma_{xx}(T)$ for our MgO/CoFeCrGa film are quite close to those reported on the same film with 12 nm thickness[60].

Furthermore, the linear temperature coefficient of the resistivity for our MgO/CoFeCrGa film was found to be $\approx -1.37 \times 10^{-10} \Omega$ m/K, which is of the same magnitude to that reported for different Heusler alloys-based spin gapless semiconductors (SGSs), such as $Mn_2CoAl$ ($-1.4 \times 10^{-9} \Omega$ m/K),[47] CoFeMnSi ($-7 \times 10^{-10} \Omega$ m/K),[61] CoFeCrAl ($-5 \times 10^{-9} \Omega$ m/K),[62] and CoFeCrGa ($-1.9 \times 10^{-9} \Omega$ m/K)[60] thin films.

We have also performed radio frequency (RF) transverse susceptibility (TS) measurements on our MgO/CoFeCrGa film in the temperature range: 20 K $\leq T \leq$ 300 K to determine the temperature evolution of effective magnetic anisotropy. This technique can accurately determine the dynamical magnetic response of a magnetic material in presence of a DC magnetic field ($H_{DC}$) and a transverse RF magnetic field ($H_{RF}$) with small and fixed amplitude.[63] When $H_{DC}$ is scanned from positive to negative saturations, the TS of a magnetic material with uniaxial anisotropy demonstrates well-defined peaks at the anisotropy fields, $H_{DC} = \pm H_K$.[64] But for a magnetic material comprising of randomly dispersed magnetic easy axes, the TS shows broad maxima at the effective anisotropy fields, $H_{DC} = \pm H_K^{eff}$. Here, we show the TS spectra as percentage change of the measured transverse susceptibility as, $\frac{\Delta \chi_T}{\chi_T}(H_{DC}) = \frac{\chi_T(H_{DC}) - \chi_T(H_{DC}^{max})}{\chi_T(H_{DC}^{max})} \times 100\%$, where $\chi_T(H_{DC}^{max})$ is the value of $\chi_T$ at the maximum value of the applied DC magnetic field, $H_{DC}^{max}$ which is chosen in such a way that $H_{DC}^{max} \gg H_{DC}^{sat}$, where $H_{DC}^{sat}$ is the saturation magnetic field. **Fig. 3**(e) shows the bipolar field scan ($+H_{DC}^{max} \rightarrow -H_{DC}^{max} \rightarrow +H_{DC}^{max}$) of $\frac{\Delta \chi_T}{\chi_T}(H_{DC})$ for MgO/CoFeCrGa film measured at $T = 20$ K for both IP ($H_{DC}$ is parallel to the film surface) and OOP ($H_{DC}$ is perpendicular to the film surface) configurations. For both the configurations, the TS shows maxima centering at $H_{DC} = \pm H_K^{eff}$. Here, we define $H_K^{eff} = H_K^{IP}$ as the IP effective anisotropy field (for IP configuration) and $H_K^{eff} = H_K^{OOP}$ as the OOP effective anisotropy field (for OOP configuration). We found that $|H_K^{OOP}| > |H_K^{IP}|$ at all the temperatures indicating IP easy axis of this film in the temperature range: 20 K $\leq T \leq$ 300 K. Furthermore, it is evident that the peaks at $+H_K^{IP}(+H_K^{OOP})$ and $-H_K^{IP}(-H_K^{OOP})$ are asymmetric with unequal peak heights which is indicative of significant anisotropy dispersion in our MgO/CoFeCrGa film for the both IP and OOP configurations. The temperature variations of $H_K^{IP}$ and $H_K^{OOP}$ for our MgO/CoFeCrGa(95nm) film are shown in **Fig. 3**(f). Clearly, both $H_K^{IP}$ and $H_K^{OOP}$ increase with decreasing temperature and $H_K^{OOP} > H_K^{IP}$ throughout the measured temperature range. Interestingly, with decreasing temperature, $H_K^{OOP}$ increases more rapidly than $H_K^{IP}$ which gives rise to large difference between $H_K^{IP}$ and $H_K^{OOP}$ at low temperatures. Additionally, both $H_K^{IP}$ and $H_K^{OOP}$ increases more rapidly below $\approx$ 200 K compared to the temperature range of 200 K $\leq T \leq$ 300 K.

### 3.3. Thermal spin transport properties: ANE and LSSE

Next, we focus on the thermo-spin transport properties of our MgO/CoFeCrGa(95nm) film. We have performed anomalous Nernst effect (ANE) and longitudinal spin Seebeck effect (LSSE) measurements on MgO/CoFeCrGa(95nm) and MgO/CoFeCrGa(95nm)/Pt(5nm) films, respectively. **Figs. 4**(a) and (b) demonstrate the schematic illustrations of our ANE and LSSE measurements. Both the ANE and LSSE measurements on MgO/CoFeCrGa(95nm) and MgO/CoFeCrGa(95nm)/Pt films, respectively were performed by sandwiching the film between two copper blocks and applying a temperature gradient (along the +z-direction) that creates a temperature difference, $\Delta T$ between those copper blocks in presence of an external DC magnetic field applied along the *x*-direction. The thermally generated Nernst and LSSE voltages generated along the *y*-direction were recorded using a Keithley 2182a nanovoltmeter while scanning the DC magnetic field. According to the theory of thermally generated magnon-driven interfacial spin pumping mechanism, simultaneous application of a vertical (*z*-axis) temperature gradient ($\overrightarrow{\nabla T}$) and an external transverse dc magnetic field ($\overrightarrow{\mu_0 H}$) (*x*-axis) across the MgO/CoFeCrGa(95nm)/Pt film gives rise to transverse spin current pumping from the CoFeCrGa layer into the Pt layer with the interfacial spin current density: $\vec{J_S} = \frac{G^{\uparrow\downarrow}}{2\pi}\frac{\gamma\hbar}{M_S V_a}k_B\vec{\nabla T}$ at the CoFeCrGa/Pt interface, where $G^{\uparrow\downarrow}$, $\hbar$, $\gamma$, $M_S$, $V_a$ and $k_B$ are the interfacial spin-mixing conductance, the reduced Planck's constant $\left(\hbar = \frac{h}{2\pi}\right)$, the gyromagnetic ratio, the saturation magnetization of CoFeCrGa, the magnon coherence volume and the Boltzmann constant, respectively[42,65,66]. The magnetic coherence volume is expressed as: $V_a = \frac{2}{3\zeta(5/2)}\left(\frac{4\pi D}{k_B T}\right)^{3/2}$ ; where, $\zeta$ is the Riemann Zeta function and $D$ is the spin-wave stiffness constant[65,66]. This transverse spin current, $\vec{J_S}$ is then converted into charge current, $\vec{J_C} = \left(\frac{2e}{\hbar}\right)\theta_{SH}^{Pt}(\vec{J_S} \times \vec{\sigma_S})$ along the y-axis via the inverse spin Hall effect (ISHE), where $e$, $\theta_{SH}^{Pt}$, and $\vec{\sigma_S}$ are the electron charge, the spin Hall angle of Pt, and the spin-polarization vector, respectively. The corresponding voltage along the y-axis can be expressed as,[42,67,68]

$$V_{LSSE} = R_y L_y \lambda_{Pt} \left(\frac{2e}{\hbar}\right)\theta_{SH}^{Pt} J_S \tanh\left(\frac{t_{Pt}}{2\lambda_{Pt}}\right), \quad (1)$$

where $R_y, L_y, \lambda_{Pt},$ and $t_{Pt}$ are the electrical resistance between the voltage leads, the distance between the voltage leads, the spin diffusion length of Pt, and the thickness of Pt layer (= 5 nm), respectively. Since CoFeCrGa is a spin-gapless semiconductor with soft ferromagnetic behavior,[51,60] concomitant application of the temperature gradient (z-axis) and dc magnetic field (x-axis) also generates a spin-polarized current in the CoFeCrGa layer along the y-axis due to ANE,[69] which gives rise to an additional contribution ($V_{CoFeCRGa}^{ANE}$) to the total voltage signal measured across the Pt layer in the MgO/CoFeCrGa(95nm)/Pt heterostructure.

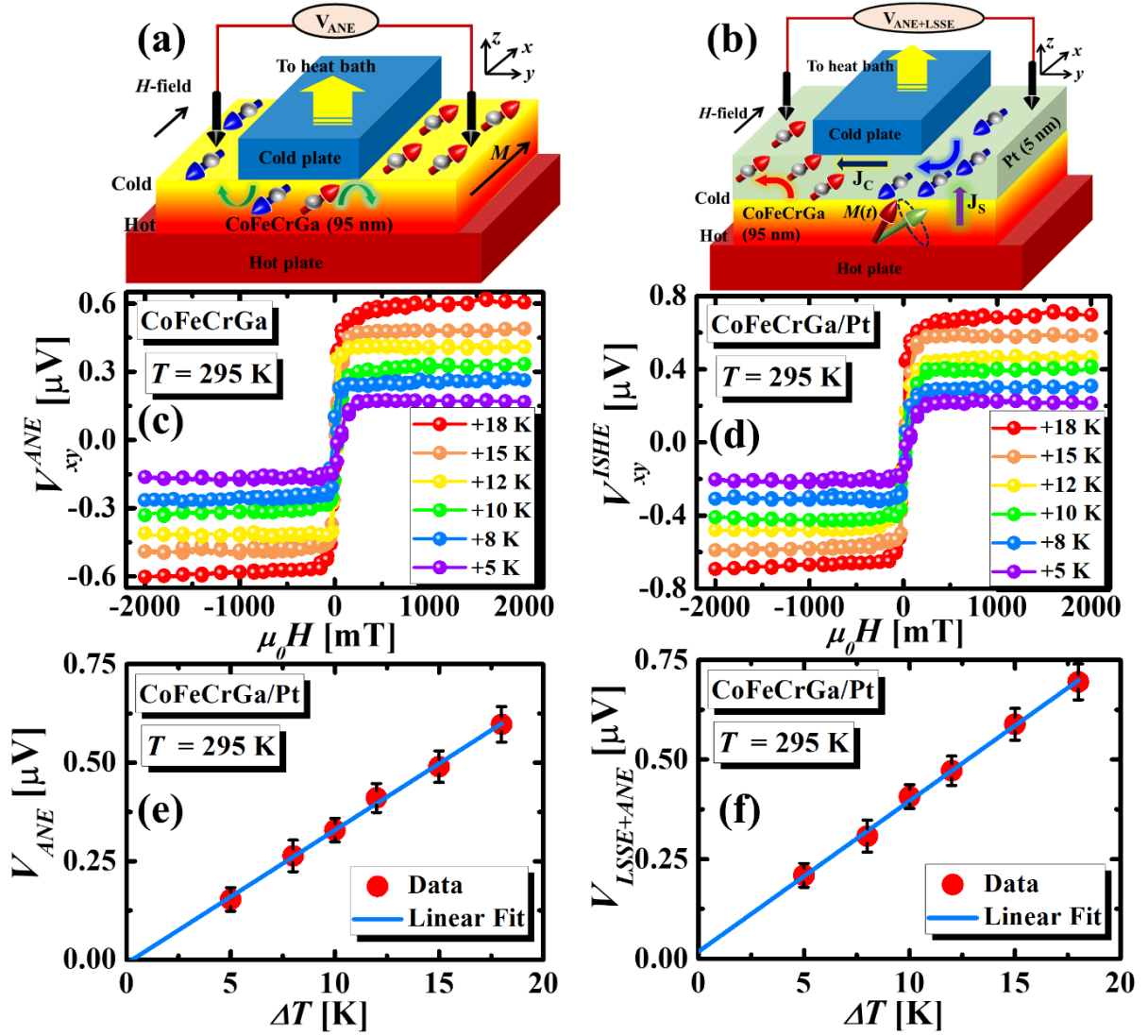

**Figure 4.** (a) and (b) the schematic illustrations of our ANE and LSSE measurements, respectively. (c) and (d) show the magnetic field dependence of the ANE voltage, $V_{ANE}(H)$ and ISHE-induced in-plane voltage, $V_{ISHE}(H)$ measured on the MgO/CoFeCrGa(95nm) and MgO/CoFeCrGa(95nm)/Pt films, respectively for different values of the temperature difference between the hot ($T_{hot}$) and cold ($T_{cold}$) copper blocks, $\Delta T = (T_{hot} - T_{cold})$ in the range: $+5\,\text{K} \leq \Delta T \leq +18\,\text{K}$ at a fixed average sample temperature $T = \frac{T_{hot}+T_{cold}}{2} = 295$ K. (e) and (f) exhibit the $\Delta T$ dependence of the background-corrected ANE voltage, $V_{ANE}(\Delta T) = \left[\frac{V_{ANE}(+\mu_0 H_{max},\ \Delta T) - V_{ANE}(-\mu_0 H_{max},\ \Delta T)}{2}\right]$ and the background-corrected (ANE+LSSE) voltage, $V_{ANE+LSSE}(\Delta T) = \left[\frac{V_{ISHE}(+\mu_0 H_{max},\ \Delta T) - V_{ISHE}(-\mu_0 H_{max},\ \Delta T)}{2}\right]$, respectively.

In presence of a transverse temperature gradient $(\vec{\nabla T})$, the electric field generated by ANE in a magnetic conductor/semiconductor with magnetization $\vec{M}$ can be expressed as,[15]

$$\vec{E_{ANE}} \propto S_{ANE}(\mu_0 \vec{M} \times \vec{\nabla T}) \qquad (2)$$

where, $S_{ANE}$ is the anomalous Nernst coefficient. Furthermore, an additional voltage contribution ($V_{Prox}^{ANE}$) can appear due to the magnetic proximity effect (MPE) induced ANE in the non-magnetic Pt layer.[69,70] Note that, only a few layers of Pt close to the CoFeCrGa(95nm)/Pt interface gets magnetized (proximitized) due to the MPE, whereas the remaining layers remain unmagnetized. Hence both $V_{CoFeCRGa}^{ANE}$ and $V_{Prox}^{ANE}$ are suppressed due to the inclusion of the 5 nm thick Pt layer on the top of CoFeCrGa layer.[41] Therefore, the resultant voltage measured across the Pt layer of our MgO/CoFeCrGa(95nm)/Pt heterostructure can be expressed as,[71] $V_{ANE+LSSE} = V_{LSSE} + V_{CoFeCRGa,\ Sup}^{ANE} + V_{Prox,\ Sup}^{ANE}$; where, $V_{CoFeCRGa,\ Sup}^{ANE}$ and $V_{Prox,\ Sup}^{ANE}$ account for the suppressed ANE voltages due to the CoFeCrGa layer and the MPE-induced ANE voltage in the Pt layer, respectively. Previous studies show that the contribution from the MPE-induced ANE in the Pt layer is negligibly small for bilayers consisting of magnetic semiconductors and Pt.[41,69] Also, in our previous report,[71] we have shown that the MPE- induced LSSE contribution of the proximitized Pt layer is negligible as only a few layers of Pt close to the CoFeCrGa(95nm)/Pt interface are magnetized due to the MPE[41]. Therefore, the resultant voltage measured across the Pt layer of our MgO/CoFeCrGa(95nm)/Pt heterostructure can be expressed as: $V_{ANE+LSSE} = V_{LSSE} + V_{CoFeCrGa,\ Sup}^{ANE}$. Considering a parallel circuit configuration of CoFeCrGa and Pt layers, the suppressed ANE voltage (due to the CoFeCrGa layer) across the Pt layer of the MgO/CoFeCrGa(95nm)/Pt heterostructure can be expressed as,[41,69]

$$V_{CoFeCRGa,\ Sup}^{ANE} = \left(\frac{F}{1+F}\right) V_{CoFeCrGa}^{ANE} \qquad (3)$$

where, $F = \frac{\rho_{Pt}}{\rho_{CoFeCrGa}} \cdot \frac{t_{CoFeCrGa}}{t_{Pt}}$, $\rho_{CoFeCrGa}$ ($\rho_{Pt}$) is the electrical resistivity of the CoFeCrGa (Pt) layer, and $t_{CoFeCrGa}$ ($t_{Pt}$) is the thickness of the CoFeCrGa (Pt) layer, respectively. Therefore, the intrinsic LSSE voltage contribution can be disentangled from the ANE contribution using the expression,[41,71]

$$V_{LSSE} = V_{ANE+LSSE} - \left(\frac{F}{1+F}\right) V_{CoFeCrGa}^{ANE} \qquad (4)$$

**Figs. 4**(c) and (d) show the magnetic field dependence of the ANE voltage, $V_{ANE}(H)$ and ISHE-induced in-plane voltage, $V_{ISHE}(H)$ measured on the MgO/CoFeCrGa(95nm) and MgO/CoFeCrGa(95nm)/Pt films, respectively for different values of the temperature difference between the hot ($T_{hot}$) and cold ($T_{cold}$) copper blocks, $\Delta T = (T_{hot} - T_{cold})$ in the range: $+5\text{ K} \leq \Delta T \leq +18\text{ K}$ at a fixed average sample temperature $T = \frac{T_{hot}+T_{cold}}{2} = 295$ K. Clearly, both $V_{ANE}(H)$ and $V_{ISHE}(H)$ signals increase upon increasing $\Delta T$. **Figs. 4**(e) and (f) exhibit the $\Delta T$ dependence of the background-corrected ANE voltage, $V_{ANE}(\Delta T) = \left[\frac{V_{ANE}(+\mu_0 H_{max},\ \Delta T) - V_{ANE}(-\mu_0 H_{max},\ \Delta T)}{2}\right]$ and the background-corrected (ANE+LSSE) voltage, $V_{ANE+LSSE}(\Delta T) = \left[\frac{V_{ISHE}(+\mu_0 H_{max},\ \Delta T) - V_{ISHE}(-\mu_0 H_{max},\ \Delta T)}{2}\right]$, respectively, where $\mu_0 H_{max}$ ($\mu_0 H_{max} \gg \mu_0 H_{sat}$) is the maximum value of the applied magnetic field strength and $\mu_0 H_{sat}$ = saturation magnetic field. Evidently, both $V_{ANE}$ and $V_{ANE+LSSE}$ scale linearly with $\Delta T$ and $|V_{ANE+LSSE}| > |V_{ANE}|$, which confirm that the observed field dependences originate from the ANE and (ANE+LSSE), respectively[41,71].

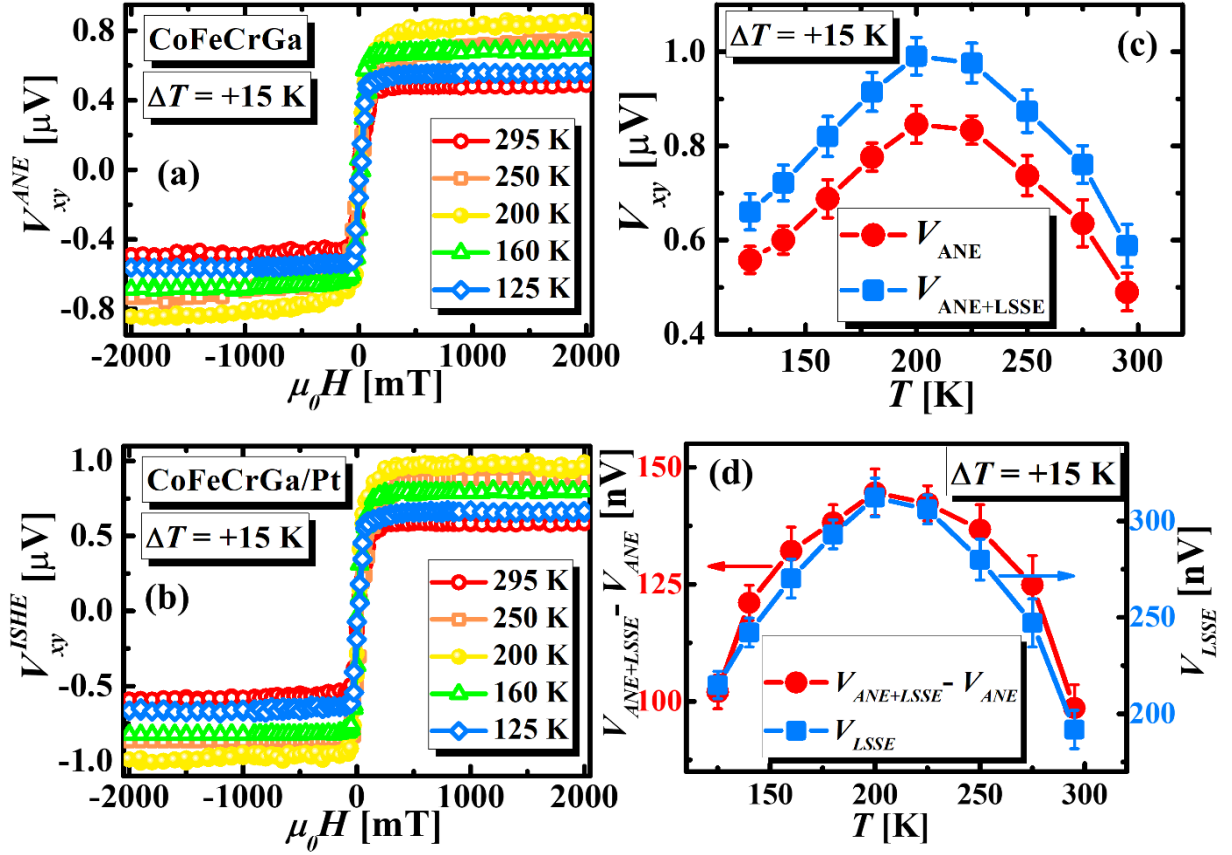

**Figure 5.** (a) and (b) $V_{ANE}(H)$ and $V_{ISHE}(H)$ hysteresis loops measured at selected average sample temperatures in the temperature range: 125 K $\leq \Delta T \leq$ 295 K for a fixed value of $\Delta T$ = +15 K on MgO/CoFeCrGa(95nm) and MgO/CoFeCrGa(95nm)/Pt films, respectively. (c) The $T$-dependence of the background-corrected ANE voltage, $V_{ANE}(T) = \left[\frac{V_{ANE}(+\mu_0 H_{max},\ T) - V_{ANE}(-\mu_0 H_{max},\ T)}{2}\right]$ and the background-corrected (ANE+LSSE) voltage, $V_{ANE+LSSE}(T) = \left[\frac{V_{ISHE}(+\mu_0 H_{max},\ T) - V_{ISHE}(-\mu_0 H_{max},\ T)}{2}\right]$ measured on MgO/CoFeCrGa(95nm) and MgO/CoFeCrGa(95nm)/Pt films, respectively. (d) Right y-scale: the temperature dependence of the intrinsic LSSE voltage, $V_{LSSE}(T)$ and the left y-scale: the temperature dependence of $[V_{ANE+LSSE}(T) - V_{ANE}(T)]$.

In **Figs. 5**(a) and (b), we show $V_{ANE}(H)$ and $V_{ISHE}(H)$ hysteresis loops measured at selected average sample temperatures in the temperature range: 125 K $\leq \Delta T \leq$ 295 K for a fixed value of $\Delta T$ = +15 K on MgO/CoFeCrGa(95nm) and MgO/CoFeCrGa(95nm)/Pt films, respectively. **Fig. 5**(c) exhibits the $T$-dependence of the background-corrected ANE voltage, $V_{ANE}(T) = \left[\frac{V_{ANE}(+\mu_0 H_{max}, T) - V_{ANE}(-\mu_0 H_{max}, T)}{2}\right]$ and the background-corrected (ANE+LSSE) voltage, $V_{ANE+LSSE}(T) = \left[\frac{V_{ISHE}(+\mu_0 H_{max}, T) - V_{ISHE}(-\mu_0 H_{max}, T)}{2}\right]$ measured on MgO/CoFeCrGa(95nm) and MgO/CoFeCrGa(95nm)/Pt films, respectively. It is evident that both $|V_{ANE}(T)|$ and $|V_{ANE+LSSE}(T)|$ increase with decreasing temperature up to $T$ = 200 K below which both of them decrease gradually with further reducing the temperature, resulting in a maximum around 200 K. Furthermore, $|V_{ANE+LSSE}(T)| > |V_{ANE}(T)|$ throughout the measured temperature range, which confirms that both ANE and LSSE contribute towards the voltage measured on the MgO/CoFeCrGa(95nm)/Pt heterostructure.

In order to determine the temperature dependence of intrinsic LSSE voltage, we have disentangled the LSSE contribution from the ANE contribution using **Eqn. 4**. The right *y*-scale of **Fig. 5**(d) shows the temperature dependence of the intrinsic LSSE voltage, $V_{LSSE}(T)$ obtained by using by **Eqn. 4** incorporating the correction factor: $\left(\frac{F}{1+F}\right)$, whereas the left *y*-scale shows the temperature dependence of the voltage difference $[V_{ANE+LSSE}(T) - V_{ANE}(T)]$ without incorporating the aforementioned correction factor, for comparison. A clear distinction can be observed between $V_{LSSE}(T)$ and $[V_{ANE+LSSE}(T) - V_{ANE}(T)]$ in terms of the absolute value as well the nature of the *T*-dependence, highlighting the importance of the correction factor for accurately determining the intrinsic LSSE contribution. Evidently, $|V_{LSSE}(T)|$ increases with decreasing temperature and shows a broad maximum around 200 K below which it decreases gradually with further lowering the temperature, as shown in **Fig. 5**(d). To ensure that the observed behavior of $V_{ANE}(T)$ and $V_{LSSE}(T)$ are intrinsic to the MgO/ CoFeCrGa(95nm) and MgO/CoFeCrGa(95nm)/Pt films, respectively, we repeated the same experiments for two more CoFeCrGa films with different thicknesses, namely, $t_{CoFeCrGa}$ = 50 and 200 nm.

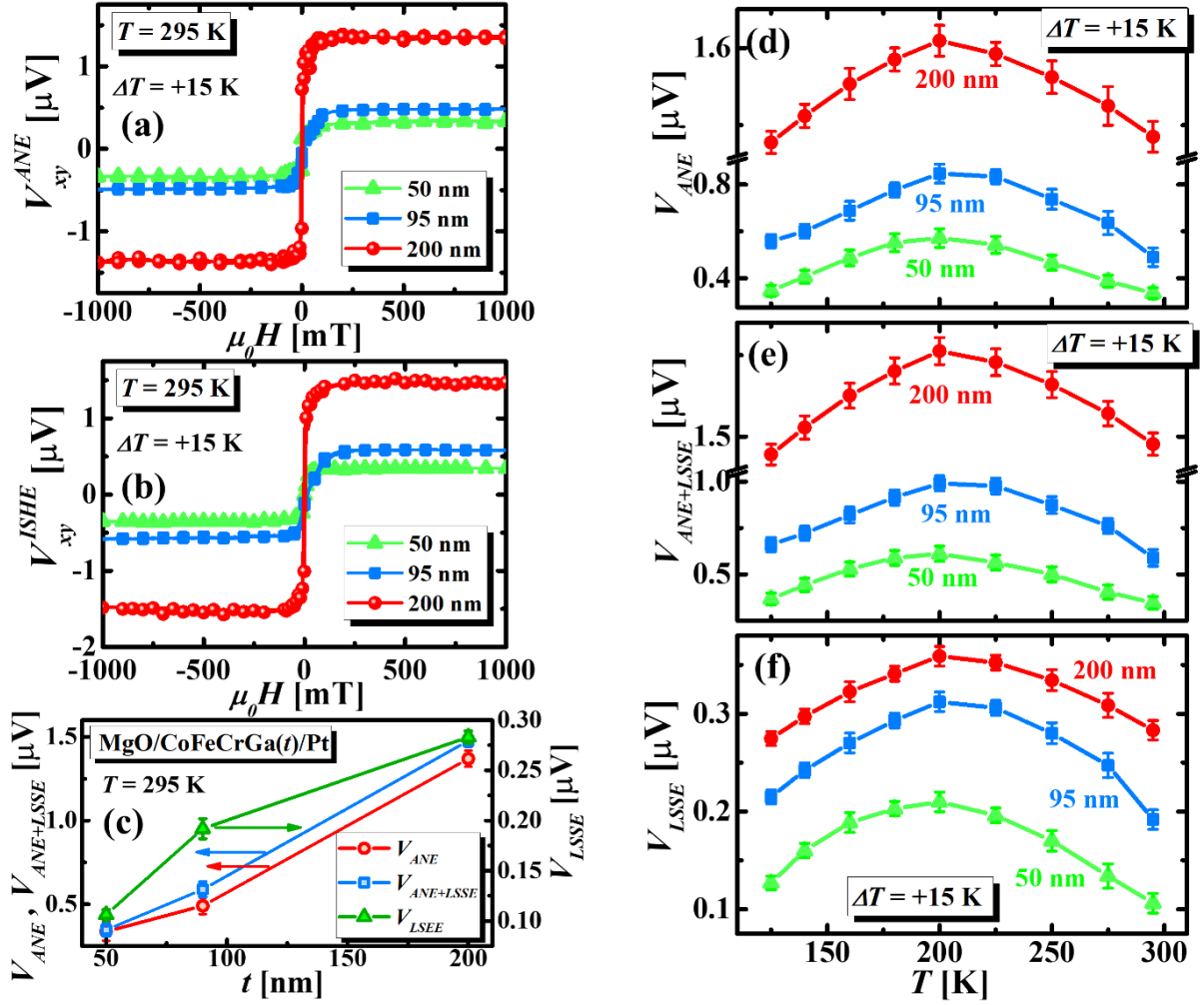

**Figure 6.** (a) and (b) $V_{ANE}(H)$ and $V_{ISHE}(H)$ hysteresis loops measured on the MgO/CoFeCrGa($t_{CoFeCrGa}$) and MgO/CoFeCrGa($t_{CoFeCrGa}$)/Pt films for $t_{CoFeCrGa}$ (CoFeCrGa film thickness) = 50, 95 and 200 nm at 295 K for $\Delta T$ = +15 K. (c) $V_{ANE}(\mu_0 H = 1\,\text{T})$, $V_{ANE+LSSE}(\mu_0 H = 1\,\text{T})$ and $V_{LSSE}(\mu_0 H = 1\,\text{T})$ at 295 K plotted as a function of $t_{CoFeCrGa}$. (d), (e) and (f) show the comparison of $V_{ANE}(T)$, $V_{ANE+LSSE}(T)$ and $V_{LSSE}(T)$, respectively for different $t_{CoFeCrGa}$.

The temperature dependent magnetometry and DC electrical transport properties of the $t_{CoFeCrGa}$ = 50 and 200 nm films are displayed in the supplementary information (**Figures S1** and **S3**). Furthermore, similar to the $t_{CoFeCrGa}$ = 95 nm film, both $V_{ANE}$ and $V_{ANE+LSSE}$ for $t_{CoFeCrGa}$ = 50 and 200 nm films scale linearly with $\Delta T$ and $|V_{ANE+LSSE}| > |V_{ANE}|$, as shown in **Figure S5**. In **Figure S6**, we demonstrate $V_{ANE}(H)$ and $V_{ISHE}(H)$ hysteresis loops at selected average sample temperatures in the range: 125 K ≤ $\Delta T$ ≤ 295 K for a fixed value of $\Delta T$ = +15 K for $t_{CoFeCrGa}$ = 50 and 200 nm films. In **Figs. 6**(a) and (b), we compare $V_{ANE}(H)$ and $V_{ISHE}(H)$ hysteresis loops measured on the MgO/CoFeCrGa($t_{CoFeCrGa}$) and MgO/CoFeCrGa($t_{CoFeCrGa}$)/Pt films for $t_{CoFeCrGa}$ (CoFeCrGa film thickness) = 50, 95 and 200 nm at $T$ = 295 K for $\Delta T$ = +15 K. As shown in **Figs. 6**(c), $V_{ANE}(\mu_0 H = 1$ T), $V_{ANE+LSSE}(\mu_0 H = 1$ T) and $V_{LSSE}(\mu_0 H = 1$ T) at 295 K increase with increasing $t_{CoFeCrGa}$. In **Figs. 6**(d), (e) and (f), we compare $V_{ANE}(T)$, $V_{ANE+LSSE}(T)$ and $V_{LSSE}(T)$, respectively for different $t_{CoFeCrGa}$. Clearly, the values of all the three quantities: $V_{ANE}$, $V_{ANE+LSSE}$ and $V_{LSSE}$ are higher for thicker CoFeCrGa films at all temperatures. Furthermore, $|V_{ANE}(T)|$, $|V_{ANE+LSSE}(T)|$ and $|V_{LSSE}(T)|$ exhibit the same behavior for all the three different CoFeCrGa film thicknesses, *i.e.*, all these quantities increase with decreasing temperature from 295 K and show a broad maximum around 200 K, which is followed by a gradual decrease with further lowering the temperature. These observations confirm that the observed behavior of $V_{ANE}(T)$ and $V_{LSSE}(T)$ are intrinsic to the MgO/CoFeCrGa and MgO/CoFeCrGa/Pt films, respectively.

### 3.4. Mechanism of LSSE and ANE at low temperatures

Since the density of the thermally generated magnons-driven spin current is proportional to the effective temperature gradient across the CoFeCrGa film through the expression, $|\vec{J_S}| = \frac{G^{\uparrow\downarrow}}{2\pi}\frac{\gamma\hbar}{M_S V_a}k_B|\vec{\nabla T}|$, it is imperative to accurately determine the effective temperature differences between the top and bottom surfaces of the CoFeCrGa film $(\Delta T_{eff})$. The total temperature difference $(\Delta T)$ across the MgO/CoFeCrGa/Pt heterostructure can be expressed as a linear combination of temperature drops in the Pt layer, at the Pt/CoFeCrGa interface, in the CoFeCrGa layer, at the CoFeCrGa/MgO interface, across the GSGG substrate as well as in the N-grease layers (thickness ≈ 1 μm) on both sides of the MgO/CoFeCrGa/Pt heterostructure, and can be written as,[72] $\Delta T = \Delta T_{Pt} + \Delta T_{\frac{Pt}{CoFeCrGa}} + \Delta T_{CoFeCrGa} + \Delta T_{\frac{CoFeCrGa}{MgO}} + \Delta T_{MgO} + 2.\Delta T_{N-Grease}$. Since the thermal resistance of Pt is very small compared to the other contributions and the bulk contributions towards the measured ISHE voltage dominate over the interfacial contributions when the thickness of the magnetic film (CoFeCrGa) is high enough,[72] the total temperature difference can be approximately written as, $\Delta T = \Delta T_{CoFeCrGa} + \Delta T_{MgO} + 2.\Delta T_{N-Grease}$.

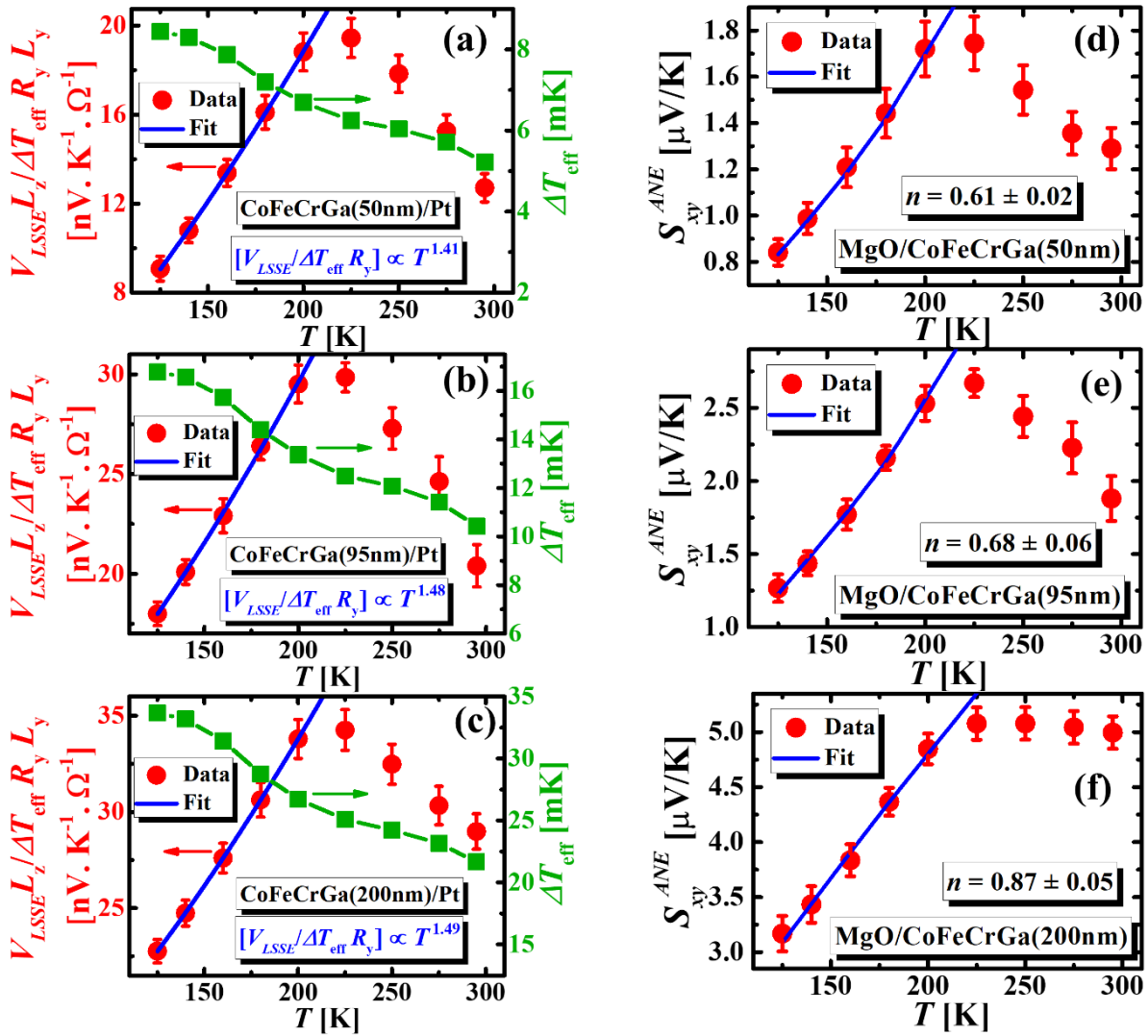

**Figure 7.** (a)-(c) Right *y*-scale: temperature dependence of $\Delta T_{eff}$ for different $t_{CoFeCrGa}$, left *y*-scale: temperature dependence of the modified LSSE coefficient, $S_{LSSE}^{eff}(T)$ for MgO/CoFeCrGa($t_{CoFeCrGa}$)/Pt(5nm) films for $t_{CoFeCrGa} = 50, 95$ and $200$ nm, respectively, fitted with the expression $S_{LSSE}^{eff} \propto \left(\theta_{SH}^{Pt} \frac{G^{\uparrow\downarrow}}{2\pi} \frac{k_B}{D^{3/2}}\right) T^n$. (d)-(f) Temperature dependence of the ANE coefficient, $S_{xy}^{ANE}(T)$ for the MgO/CoFeCrGa($t_{CoFeCrGa}$) films for $t_{CoFeCrGa} =$ 50, 95 and 200 nm, respectively, fitted with **Eqn. 6**.

Considering the 4-slab model, the total thermal resistance between hot and cold plates can be written as, $R_{Th} = \frac{1}{A}\left(\frac{2t_{N-Grease}}{\kappa_{N-Grease}} + \frac{t_{CoFeCrGa}}{\kappa_{CoFeCrGa}} + \frac{t_{MgO}}{\kappa_{MgO}}\right)$, where, $t_{N-Grease}$, $t_{MgO}$ and $t_{CoFeCrGa}$ are the thicknesses of the grease layers, MgO substrate and the CoFeCrGa layer, respectively; $\kappa_{N-Grease}$, $\kappa_{MgO}$ and $\kappa_{CoFeCrGa}$ are the thermal conductivities of the grease layers, MgO substrate and the CoFeCrGa layer, respectively, and $A$ is the cross sectional area. Since the rate of heat flow across the entire heterostructure reaches a constant value in the steady state, the effective temperature difference across the CoFeCrGa film can be written as,[42]

$$\Delta T_{eff} = \Delta T_{CoFeCrGa} = \frac{\Delta T}{\left[1 + \frac{\kappa_{CoFeCrGa}}{t_{CoFeCrGa}}\left(\frac{2t_{N-Grease}}{\kappa_{N-Grease}} + \frac{t_{MgO}}{\kappa_{MgO}}\right)\right]} \quad (5)$$

We have measured the temperature dependence of thermal conductivity of bulk CoFeCrGa using the thermal transport option of the PPMS, as shown in the supplementary information (**Figure S4**). Using the reported values of the thermal conductivities of the Apiezon N-grease,[73] and the MgO crystal[74], we have determined the temperature dependence of $\Delta T_{eff}$ for different $t_{CoFeCrGa}$ using **Eqn. 5**, as shown in **Figs. 7**(a)-(c). Here, we have ignored the interfacial thermal resistances between the N-grease and the hot/cold plates as well as between the sample and N-grease layers.[14]

Using the *T*-dependence of $\Delta T_{eff}$, we have estimated the *T*-dependence of the modified LSSE coefficient, $S_{LSSE}^{eff}(T) = \frac{V_{LSSE}(T)}{\Delta T_{eff}(T) R_y(T)} \times \left(\frac{L_z}{L_y}\right)$ for MgO/CoFeCrGa($t_{CoFeCrGa}$)/Pt(5nm) films for $t_{CoFeCrGa} = 50, 95$ and 200 nm; where, $L_y$ (= 3 mm) is the distance between the voltage leads and $L_z = t_{CoFeCrGa}$ (see **Figs. 7**(a)-(c)). Note that we have measured the *T*-dependence of resistance $\left(R_y(T)\right)$ between the voltage-leads placed on the Pt layer of the MgO/CoFeCrGa($t_{CoFeCrGa}$)/Pt heterostructures using 4-point probe configuration. Note that the value of $S_{LSSE}^{eff}(T)$ for our MgO/CoFeCrGa($t_{CoFeCrGa}$)/Pt heterostructures are $\approx$ 12.8, 20.5 and 29.8 nV.K$^{-1}$.Ω$^{-1}$ at $T$ = 295 K for $t_{CoFeCrGa}$ = 50, 95 and 200 nm, respectively, which are higher than that of the half-metallic FM thin films of La$_{0.7}$Sr$_{0.3}$MnO$_3$ ($\approx$ 9 nV.K$^{-1}$.Ω$^{-1}$ at room temperature)[43]. As shown in **Figs. 7**(a)-(c), $S_{LSSE}^{eff}(T)$ for the MgO/CoFeCrGa($t_{CoFeCrGa}$)/Pt(5nm) heterostructures for all the three CoFeCrGa film thicknesses increases as $T$ decreases from room temperature and shows a peak around 225 K below which it decreases rapidly with further decrease in temperature. Since the saturation magnetization, $M_S \approx T^{-1}$ in the measured temperature range (as shown in **Fig. 3**(a)) and, $V_a \propto T^{-3/2}$, according to the theory of magnon-driven LSSE, $|\vec{J_S}| \propto \frac{G^{\uparrow\downarrow}}{2\pi} \frac{k_B}{D^{3/2}} T^{5/2} |\vec{\nabla T}|$.[42] Considering $\tanh\left(\frac{t_{Pt}}{2\lambda_{Pt}}\right) \approx 1$ for our case and, $\lambda_{Pt} \propto T^{-1}$,[75] according to the **Eqn. 1**, the modified LSSE coefficient becomes $S_{LSSE}^{eff} = \frac{V_{LSSE}}{\Delta T_{eff} R_y L_y} \propto \left(\theta_{SH}^{Pt} \frac{G^{\uparrow\downarrow}}{2\pi} \frac{k_B}{D^{3/2}}\right) T^{3/2}$.[42] As shown in **Figs. 7**(a)-(c), $S_{LSSE}^{eff}(T)$ varies as $T^{1.41\pm0.12}$, $T^{1.48\pm0.08}$ and $T^{1.49\pm0.1}$ for $t_{CoFeCrGa}$ = 50, 95 and 200 nm, respectively in the measured temperature range, which are in good agreement with the theory of thermally generated magnon-driven interfacial spin pumping mechanism[42,65,66].

Now, let us understand the origin of ANE in our MgO/CoFeCrGa($t_{CoFeCrGa}$) films. The transverse thermoelectric coefficient ($S_{xy}$) is expressed as, $S_{xy} = \left[\frac{\alpha_{xy} - S_{xx}\sigma_{xy}}{\sigma_{xx}}\right]$, where $\sigma_{xx}$ and $\sigma_{xy}$ are the longitudinal and transverse electrical conductivities which are defined as,[9,18,22] $\sigma_{xx} = \left[\frac{\rho_{xx}}{(\rho_{xx})^2 + (\rho_{xy})^2}\right]$ and $\sigma_{xy} = \left[\frac{-\rho_{xy}}{(\rho_{xx})^2 + (\rho_{xy})^2}\right]$, respectively. Also, $\alpha_{xy}$ and $S_{xx}$ are the transverse thermoelectric conductivity and longitudinal Seebeck coefficient, which according to the Mott's relations can be expressed as, [9,15,76] $\alpha_{xy} = \frac{\pi^2 k_B^2 T}{3e}\left(\frac{\partial \sigma_{xy}}{\partial E}\right)_{E=E_F}$ and $S_{xx} = \frac{\pi^2 k_B^2 T}{3e\sigma_{xx}}\left(\frac{\partial \sigma_{xx}}{\partial E}\right)_{E=E_F}$, respectively, where $E_F$ is the Fermi energy. Since ANE and anomalous Hall Effect (AHE) share the common physical origin and the AHE follows the power law connecting the anomalous Hall resistivity, $\rho_{xy}^{AHE}$ with the longitudinal electrical resistivity, $\rho_{xx}$ through the expression, $\rho_{xy}^{AHE} = \lambda M \rho_{xx}^n$,[9] where $\lambda$ is the spin-orbit coupling constant and $n$ is a constant exponent, the anomalous Nernst coefficient can be expressed as,[9,15]

$$S_{xy}^{ANE} = \rho_{xx}^{n-1}\left[\frac{\pi^2 k_B^2 T}{3e}\left(\frac{\partial \lambda}{\partial E}\right)_{E=E_F} - (n-1)\lambda S_{xx}\right]. \tag{6}$$

When $n = 1$, the extrinsic skew scattering is the predominant mechanism for the anomalous Nernst/Hall transport, whereas $n = 2$ indicates the intrinsic Berry curvature or, the extrinsic side jump dominated anomalous Nernst/Hall transport[25]. Using the T-dependences of ANE voltage, $V_{ANE}(T)$ and $\Delta T_{eff}$, we have estimated the $T$-dependence of the ANE coefficient, $S_{xy}^{ANE}(T) = \frac{V_{ANE}(T)}{\Delta T_{eff}(T)} \times \left(\frac{L_z}{L_y}\right)$ for the MgO/CoFeCrGa($t_{CoFeCrGa}$) films, as shown in **Figs. 7**(d)-(f). Similar to the modified LSSE voltage, $V_{LSSE}^{eff}(T)$, $S_{xy}^{ANE}(T)$ for the MgO/CoFeCrGa($t_{CoFeCrGa}$) films for all the three CoFeCrGa film thicknesses also increases as $T$ decreases from room temperature and shows a maximum around 225 K below which it decreases rapidly with further decrease in temperature. Interestingly, $S_{xy}^{ANE}(T)$ for the MgO/CoFeCrGa(200 nm) film increases slowly with decreasing temperature from the room temperature and the maximum around 225 K is much broader in contrast to the films with lower thicknesses.

Note that the values of $S_{xy}^{ANE}$ for our MgO/CoFeCrGa($t_{CoFeCrGa}$) films are $\approx$ 1.28, 1.86 and 4.9 $\mu V.K^{-1}$ at $T = 295$ K and $\approx$ 1.75, 2.63 and 5.1 $\mu V.K^{-1}$ at 225 K, for $t_{CoFeCrGa} = 50, 95$ and 200 nm, respectively which are nearly two orders of magnitude higher than that of the bulk polycrystalline sample of CoFeCrGa ($\approx 0.018 \, \mu V. K^{-1}$ at 300 K)[14] but, comparable to that of the magnetic Weyl semimetal $Co_2MnGa$ thin films ($\approx 2 - 3 \, \mu V. K^{-1}$ at 300 K)[77,78]. We fitted the $S_{xy}^{ANE}(T)$ data in the temperature range 125 K $\leq T \leq$ 200 K for our MgO/CoFeCrGa($t_{CoFeCrGa}$) films using **Eqn. 6** considering $\lambda$, $\left(\frac{\partial \lambda}{\partial E}\right)_{E=E_F}$, and $n$ as the fitting parameters. The best fit was obtained for $n = 0.61 \pm 0.02$, $0.68 \pm 0.06$ and $0.87 \pm 0.05$, for $t_{CoFeCrGa} = 50, 95$ and 200 nm, respectively which implies that the origin of ANE in our MgO/CoFeCrGa($t_{CoFeCrGa}$) films is dominated by the asymmetric skew scattering of charge carriers below 200 K[25]. Note that, we have also observed skew-scattering dominated ANE in bulk polycrystalline sample of CoFeCrGa,[14] for which $n \approx 0.78$.

Next, let us examine the temperature evolution of the anomalous off-diagonal thermoelectric conductivity, $\alpha_{xy}^{ANE}(T)$. To determine $\alpha_{xy}^{ANE}(T)$, we have performed the Hall measurements on the MgO/CoFeCrGa($t_{CoFeCrGa}$) films. **Figs. 8**(a)-(c) present the magnetic field dependence of Hall resistivity $\rho_{xy}(H)$ of our MgO/CoFeCrGa($t_{CoFeCrGa}$) films for $t_{CoFeCrGa} = 50, 95$ and 200 nm, respectively recorded at few selected temperatures in the range: 125 K $\leq T \leq$ 295 K. By subtracting the ordinary Hall effect (OHE) contribution from $\rho_{xy}(H)$, we determined the *T*-dependence of the anomalous Hall resistivity $\rho_{xy}^{AHE}(T)$. The left-*y* scales of **Figs. 8**(d)-(f) exhibit the T-dependence of the anomalous Hall conductivity, $\left|\sigma_{xy}^{AHE}\right| = \left[\frac{\rho_{xy}^{AHE}}{(\rho_{xx})^2 + (\rho_{xy}^{AHE})^2}\right]$ of our MgO/CoFeCrGa($t_{CoFeCrGa}$) films for $t_{CoFeCrGa} = 50, 95$ and 200 nm, respectively. Note that $\left|\sigma_{xy}^{AHE}(T)\right|$ for our MgO/CoFeCrGa($t_{CoFeCrGa}$) films increases almost linearly with decreasing temperature, unlike $UCo_{0.8}Ru_{0.2}Al$ for which $\left|\sigma_{xy}^{AHE}\right|$ is nearly temperature independent at low temperatures[79]. This implies that $\sigma_{xy}^{AHE}$ for our MgO/CoFeCrGa(95nm) film is strongly dependent on the scattering rate, which further supports that the extrinsic mechanisms (*e.g.*, asymmetric skew scattering) dominate the transverse thermoelectric response of our sample at low temperatures[79].

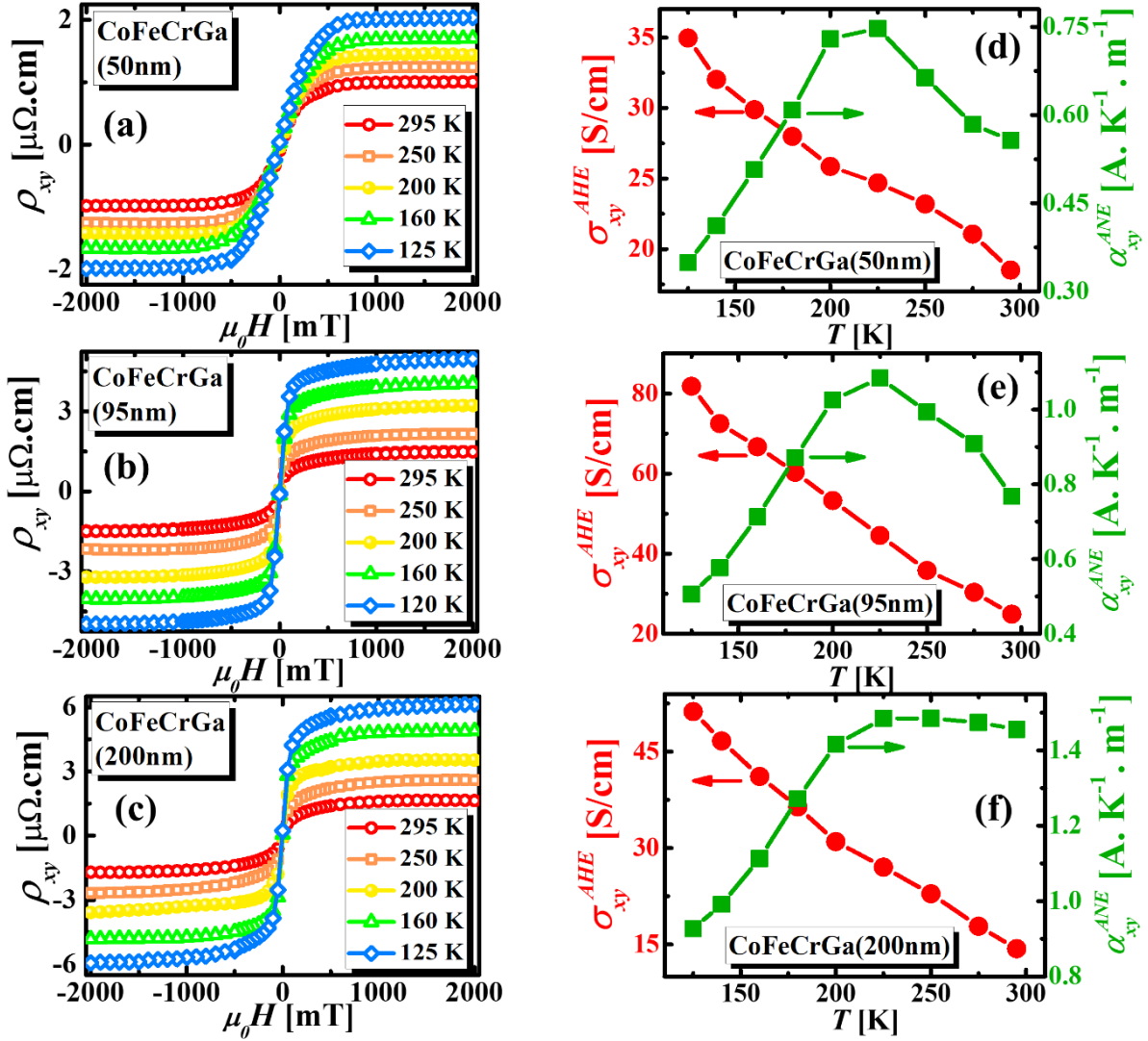

**Figure 8.** (a)-(c) Magnetic field dependence of Hall resistivity $\rho_{xy}(H)$ of our MgO/ CoFeCrGa($t_{CoFeCrGa}$) films for $t_{CoFeCrGa} = 50, 95$ and $200$ nm, respectively recorded at few selected temperatures in the range: $125\ \text{K} \leq T \leq 295\ \text{K}$. (d)-(f) Left y-scale: the temperature dependence of the anomalous Hall conductivity, $\left|\sigma_{xy}^{AHE}\right|$ of our MgO/CoFeCrGa($t_{CoFeCrGa}$) films for $t_{CoFeCrGa} = 50, 95$ and $200$ nm, respectively, right y-scale: corresponding temperature variations of transverse thermoelectric conductivity $\alpha_{xy}^{ANE}$.

The right $y$-scale of **Figs. 8**(d)-(f) illustrates the temperature variation of $\alpha_{xy}^{ANE}$ of our MgO/CoFeCrGa($t_{\text{CoFeCrGa}}$) films for $t_{\text{CoFeCrGa}} = 50, 95$ and $200$ nm, respectively, which was obtained by incorporating the $T$-dependences of $S_{xx}$, $S_{ANE}$, $\rho_{xx}$ and $\rho_{xy}^{AHE}$ in the expression,[20,21,80] $\alpha_{xy}^{ANE} = S_{xy}^{ANE}\sigma_{xx} + S_{xx}\sigma_{xy}^{AHE} = \left[\frac{S_{xy}^{ANE}\rho_{xx} - S_{xx}\rho_{xy}^{AHE}}{(\rho_{xx})^2 + (\rho_{xy}^{AHE})^2}\right]$. It is evident that $\alpha_{xy}^{ANE}(T)$ for all the films shows a maximum around 225 K, similar to $S_{xy}^{ANE}(T)$. Note that similar to $S_{xy}^{ANE}(T)$, $\alpha_{xy}^{ANE}(T)$ for the MgO/CoFeCrGa(200 nm) film increases slowly with decreasing temperature from the room temperature and the maximum around 225 K is much broader in contrast to the films with lower thicknesses. The values of $\alpha_{xy}^{ANE}$ at room temperature (295 K) for our MgO/CoFeCrGa($t_{\text{CoFeCrGa}}$) films are $0.55, 0.77$ and $1.4$ A.m$^{-1}$.K$^{-1}$ for $t_{\text{CoFeCrGa}} = 50, 95$ and $200$ nm, respectively, which are much smaller than that of non-centrosymmetric Kagome ferromagnet UCo$_{0.8}$Ru$_{0.2}$Al[79] ($\approx 15$ A.m$^{-1}$.K$^{-1}$ at 40 K), Co$_2$MnGa single crystal[18] ($\approx 7$ A.m$^{-1}$.K$^{-1}$ at 300 K) but closer to that of Co$_2$MnGa thin films[78] ($\approx 2$ A.m$^{-1}$.K$^{-1}$ at 300 K).

Next, we focus on the origin of the maximum in both $S_{LSSE}^{eff}(T)$ and $S_{xy}^{ANE}(T)$ centered around 225 K. Note that the occurrence of maximum in both LSSE and ANE signals at the same temperature has been observed in other ferromagnetic metallic films, *e.g.*, mixed valent manganites[42], iron oxides[71]. The maximum in the temperature dependent LSSE signal in the magnetically ordered state is commonly observed in different ferro- and ferrimagnets for example, YIG, La$_{0.7}$Ca$_{0.3}$MnO$_3$ etc., which originates as a consequence of the combined effects of boundary scattering and diffusive inelastic magnon-phonon or magnon-magnon scattering processes together with the reduction of magnon population at low temperatures[33,42,81]. In YIG, the maximum in the LSSE signal is thickness dependent; it shifts from $\approx 70$ K for bulk YIG slab to $\approx 200$ K for 1 μm YIG film.[33] In ferromagnetic metals, extrinsic contributions arising from electron-magnon scattering contributes significantly to the anomalous Nernst thermopower.[45] In presence of a temperature gradient and external magnetic field, magnons are excited in the bulk of a ferromagnetic material and these thermally generated magnons transfer spin-angular momenta to the itinerant electrons via electron-magnon scattering as a result of which the itinerant electrons of the ferromagnetic layer get spin polarized and contribute to the ANE.[45] Since the observed ANE in our MgO/CoFeCrGa($t_{CoFeCrGa}$) films has dominating contribution from the extrinsic mechanism, the occurrence of maxima in $S_{xy}^{ANE}(T)$ around 225 K and the subsequent decrease in $S_{xy}^{ANE}$ in our MgO/CoFeCrGa($t_{CoFeCrGa}$) films can also be attributed to the diffusive inelastic magnon scatterings and reduced magnon population at low temperatures.[45] A decrease in the magnon population at low temperatures also reduces electron-magnon scattering which eventually diminishes the population of the spin-polarized itinerant electrons participating in the skew-scattering process.

In case of LSSE, the magnon propagation length ($\langle \xi \rangle$) of the ferromagnetic material also plays vital role in addition to the magnon population. $\langle \xi \rangle$ signifies the critical length scale for thermally-generated magnons to develop a spatial gradient of magnon accumulation inside a ferromagnetic film which is one of the crucial factors that governs spin angular momentum transfer to the adjacent HM layer[31,33,82]. A decrease in $\langle \xi \rangle$ also suppresses the LSSE signal. It was theoretically shown that $\langle \xi \rangle$ of a magnetic material with lattice constant $a_0$ is related to the effective anisotropy constant ($K_{eff}$) and the Gilbert damping parameter ($\alpha$), through the relation[34,82] $\langle \xi \rangle = \frac{a_0}{2\alpha} \cdot \sqrt{\frac{J_{ex}}{2K_{eff}}}$, , where $J_{ex}$ is the strength of the Heisenberg exchange interaction between nearest neighbors. Since $K_{eff} = \frac{1}{2} M_S H_K^{eff}$, the aforementioned expression can be written as, $\langle \xi \rangle = \frac{a_0}{2\alpha} \cdot \sqrt{\frac{J_{ex}}{\mu_0 M_S H_K^{eff}}}$. Thus, $\langle \xi \rangle$ is inversely proportional to $\alpha$ as well as the square-root of $(M_S H_K^{eff})$. This implies that the $T$-evolution of $\langle \xi \rangle$ is related to that of $\alpha$, $H_K^{eff}$ and $M_S$. As shown in **Fig. 3**(a), $M_S$ for our MgO/CoFeCrGa(95nm) film increases with decreasing temperature. Furthermore, $H_K^{eff}$ of our MgO/CoFeCrGa(95nm) film for both IP and OOP configurations (both $H_K^{IP}$ and $H_K^{OOP}$) increases with decreasing temperature and the increase is more rapid below $\approx$ 200 K compared to the temperature range of 200 K $\leq T \leq$ 300 K, as indicated in **Fig. 3**(f). Notably, similar behavior of $H_K^{eff}$ has been observed for the MgO/CoFeCrGa(200nm) film (see **Figure S2**). Therefore, both $H_K^{eff}$ and $M_S$ tends to suppress $\langle \xi \rangle$ (and hence, $S_{LSSE}^{eff}$) at low temperatures, especially below $\approx$ 200 K. To comprehend the role of $\alpha$ in $\langle \xi \rangle$ and hence, the LSSE signal at low temperatures, we have investigated the spin-dynamic properties of our MgO/CoFeCrGa(95nm) and MgO/CoFeCrGa(95nm)/Pt(5nm) films by employing the broadband ferromagnetic resonance (FMR) measurements.

### 3.5. Magnetization dynamics and Gilbert damping

**Figs. 9**(a) and (b) display the field-derivative of microwave (MW) power absorption spectra $\left(\frac{dP}{dH}\right)$ as a function of the IP DC magnetic field for various frequencies in the range: 4 GHz $\leq f \leq$ 18 GHz recorded at $T$ = 250 K for the MgO/CoFeCrGa(95nm) and MgO/CoFeCrGa(95nm)/Pt(5nm) films, respectively. To extract the resonance field ($H_{res}$) and linewidth ($\Delta H$), we fitted the $\frac{dP}{dH}$ lineshapes with a linear combination of symmetric and antisymmetric Lorentzian function derivatives as,[83]

$$\frac{dP}{dH} = P_{Sym} \frac{\frac{\Delta H}{2}(H_{dc}-H_{res})}{\left[(H_{dc}-H_{res})^2+\left(\frac{\Delta H}{2}\right)^2\right]^2} + P_{Asym} \frac{\left(\frac{\Delta H}{2}\right)^2-(H_{dc}-H_{res})^2}{\left[(H_{dc}-H_{res})^2+\left(\frac{\Delta H}{2}\right)^2\right]^2} + P_0 \qquad (7)$$

where, $P_{Sym}$ and $P_{Asym}$ are the coefficients of the symmetric and antisymmetric Lorentzian derivatives, and $P_0$ is a constant offset parameter. The fitted curves are represented by solid lines in **Figs. 9**(a) and (b). To obtain the temperature evolution of the damping parameter, $\alpha(T)$ for the MgO/CoFeCrGa(95nm) and MgO/CoFeCrGa(95nm)/Pt(5nm) films, we have fitted the $\Delta H$-$f$ curves with the expression,[84] $\Delta H = \Delta H_0 + \frac{4\pi\alpha}{\gamma\mu_0}f$, where, $\Delta H_0$ represents the inhomogeneous broadening, $\frac{\gamma}{2\pi} = \frac{g_{eff}\mu_B}{\hbar}$ is the gyromagnetic ratio, $\mu_B$ is the Bohr magneton, $g_{eff}$ is the Landé g-factor. **Figs. 9**(c) shows the $\Delta H$-$f$ curves for the MgO/CoFeCrGa(95nm) film at different temperatures fitted with the aforementioned expression. Clearly, the slope of the $\Delta H$-$f$ curves increases with decreasing temperature which implies increase $\alpha$ at low temperatures. In **Fig. 9**(d), we compare the $\Delta H$-$f$ curves for the MgO/CoFeCrGa(95nm) and MgO/CoFeCrGa(95nm)/Pt(5nm) films recorded at $T$ = 250 K. It is evident that $\Delta H$ for MgO/CoFeCrGa(95nm)/Pt(5nm) is higher than that of MgO/CoFeCrGa(95nm) for all the frequencies, which is because of the loss of spin angular momentum in the CoFeCrGa film as a result of spin pumping and can be expressed as,[85] $[\Delta H_{CoFeCrGa/Pt} - \Delta H_{CoFeCrGa}] = G_R^{\uparrow\downarrow}\left(\frac{g_{eff}\mu_B}{2\gamma M_s t_{CoFeCrGa}}\right)f$, where $G_R^{\uparrow\downarrow}$ is the real component of the interfacial spin mixing conductance $(G^{\uparrow\downarrow})$. From the fits, we obtained $\alpha_{CoFeCrGa} = (3.6 \pm 0.2) \times 10^{-2}$ and $\alpha_{CoFeCrGa/Pt} = (4.12 \pm 0.1) \times 10^{-2}$ at 250 K for the MgO/CoFeCrGa(95nm), and MgO/CoFeCrGa(95nm)/Pt(5nm) films, respectively. Clearly, $\alpha_{CoFeCrGa/Pt} > \alpha_{CoFeCrGa}$ which is caused by additional damping due to the spin pumping effect[85]. In **Fig. 9**(e), we compare $\alpha(T)$ for the MgO/CoFeCrGa(95nm), and MgO/CoFeCrGa(95nm)/Pt(5nm) films. It is evident that $\alpha_{CoFeCrGa/Pt} > \alpha_{CoFeCrGa}$ at all the temperatures and both $\alpha_{CoFeCrGa/Pt}$ and $\alpha_{CoFeCrGa}$ increase with decrease temperature, especially below 225 K. Such increase in $\alpha$ and $\Delta H$ at low temperatures can be primarily attributed to the impurity relaxation mechanisms[86–88]. Since $\langle\xi\rangle \propto \frac{1}{\alpha}$, an increase in $\alpha$ at low temperatures gives rise to decrease in $\langle\xi\rangle$, and hence, the LSSE signal. The increases in $\Delta H_0$ at low temperatures for both MgO/CoFeCrGa(95nm) MgO/

CoFeCrGa(95nm)/Pt(5nm) films (see right *y*-scale of **Fig. 9**(f)) also support the occurrence of impurity relaxation at low temperatures[89].

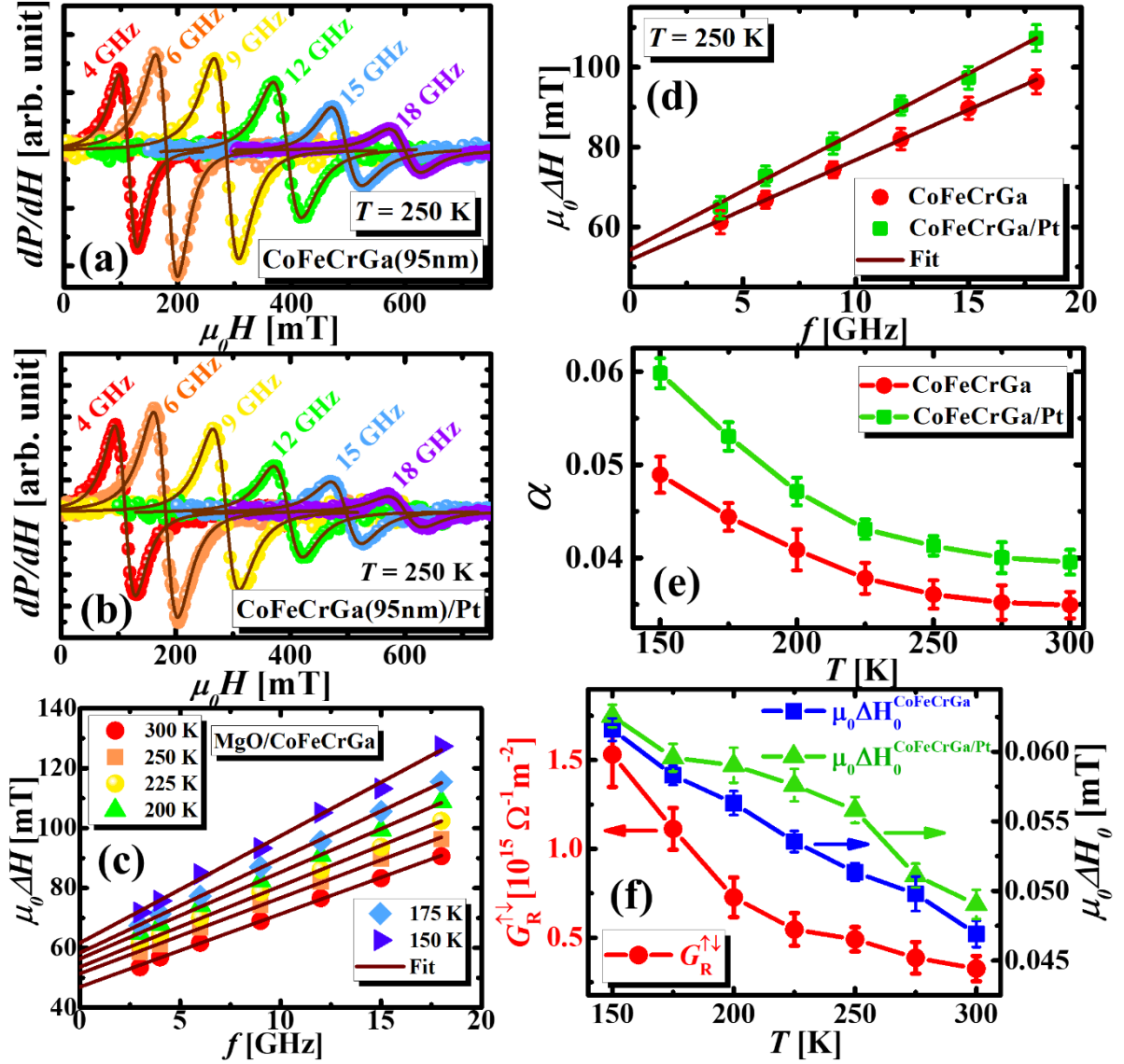

**Figure 9.** (a) and (b) Field-derivative of microwave (MW) power absorption spectra $\left(\frac{dP}{dH}\right)$ as a function of the IP DC magnetic field for various frequencies in the range: 4 GHz $\leq f \leq$ 18 GHz recorded at 250 K for the MgO/CoFeCrGa(95nm) and MgO/CoFeCrGa(95nm)/Pt(5nm) films, respectively fitted with **Eqn. 7**. (c) The $\Delta H$-$f$ curves for the MgO/CoFeCrGa(95nm) film at different temperatures fitted with $\Delta H = \Delta H_0 + \frac{4\pi\alpha}{\gamma\mu_0}f$. (d) The comparison of the $\Delta H$-$f$ curves for the MgO/CoFeCrGa(95nm) and MgO/CoFeCrGa(95nm)/Pt(5nm) films recorded at 250 K. (e) Comparison of the temperature dependence of damping parameter $\alpha(T)$ for the MgO/CoFeCrGa(95nm), and MgO/CoFeCrGa(95nm)/Pt(5nm) films. (f) Right y-scale: temperature dependence of $\Delta H_0$ for the MgO/CoFeCrGa(95nm) and MgO/CoFeCrGa(95nm)/Pt films, left y-scale: temperature dependence of the real component of the spin mixing conductance $G_R^{\uparrow\downarrow}$ for the MgO/CoFeCrGa(95nm)/Pt(5nm) film.

To have a quantitative understanding of the *T*-evolution of spin pumping efficiency in the MgO/CoFeCrGa(95nm)/Pt(5nm) film, we estimated $G_R^{\uparrow\downarrow}$ using the expression,[90] $G_R^{\uparrow\downarrow} = \left(\frac{2e^2}{h}\right)\left(\frac{2\pi M_s t_{CoFeCrGa}}{g_{eff}\mu_B}\right)[\alpha_{CoFeCrGa/Pt} - \alpha_{CoFeCrGa}]$ where, $G_0 = \left(\frac{2e^2}{h}\right)$ is the conductance quantum, and found that $G_R^{\uparrow\downarrow} \approx 3.25 \times 10^{14}\ \Omega^{-1}m^{-2}$ at 300 K which is close to $G_R^{\uparrow\downarrow} = 7.5 \times 10^{14}\ \Omega^{-1}m^{-2}$ in YIG/Pt[91] and $G_R^{\uparrow\downarrow} = 5.7 \times 10^{14}\ \Omega^{-1}m^{-2}$ in TmIG/Pt bilayers[90]. As shown in **Fig. 9**(d), $G_R^{\uparrow\downarrow}$ for the MgO/CoFeCrGa(95nm)/Pt(5nm) film increases with decreasing temperature, which is consistent with the phenomenological expression,[92] $G_R^{\uparrow\downarrow} \propto (T_C - T)$, where $T_C$ = Curie temperature. Furthermore, to confirm the aforementioned behavior of the temperature evolution of $\alpha$, we have repeated the broadband FMR measurements on the MgO/CoFeCrGa(200nm)/Pt(5nm) film. **Figs. 10**(a) display the magnetic field dependence of the $\left(\frac{dP}{dH}\right)$ lineshapes in the range: 6 GHz $\leq f \leq$ 24 GHz recorded at *T* = 250 K for the MgO/CoFeCrGa(200nm)/Pt film, fitted with **Eqn. 7**. To obtain the temperature evolution of the damping parameter, $\alpha(T)$ we have fitted the $\Delta H$-*f* curves at different temperatures in the range of 140 K $\leq T \leq$ 300 K with the expression,[84] $\Delta H = \Delta H_0 + \frac{4\pi\alpha}{\gamma\mu_0}f$, as shown in **Fig. 10**(b). Evidently, the slope of the $\Delta H$-*f* curves increases with decreasing temperature which implies increase $\alpha$ at low temperatures. Moreover, in **Fig. 10**(c), we show the fitting of the *f*-$H_{res}$ curves at *T* = 250 K using Kittel's equation for magnetic thin films with IP magnetic field,[86] which is expressed as, $f = \frac{\gamma\mu_0}{2\pi}\sqrt{H_{res}(H_{res} + M_{eff})}$, where $M_{eff}$ is the effective magnetization.

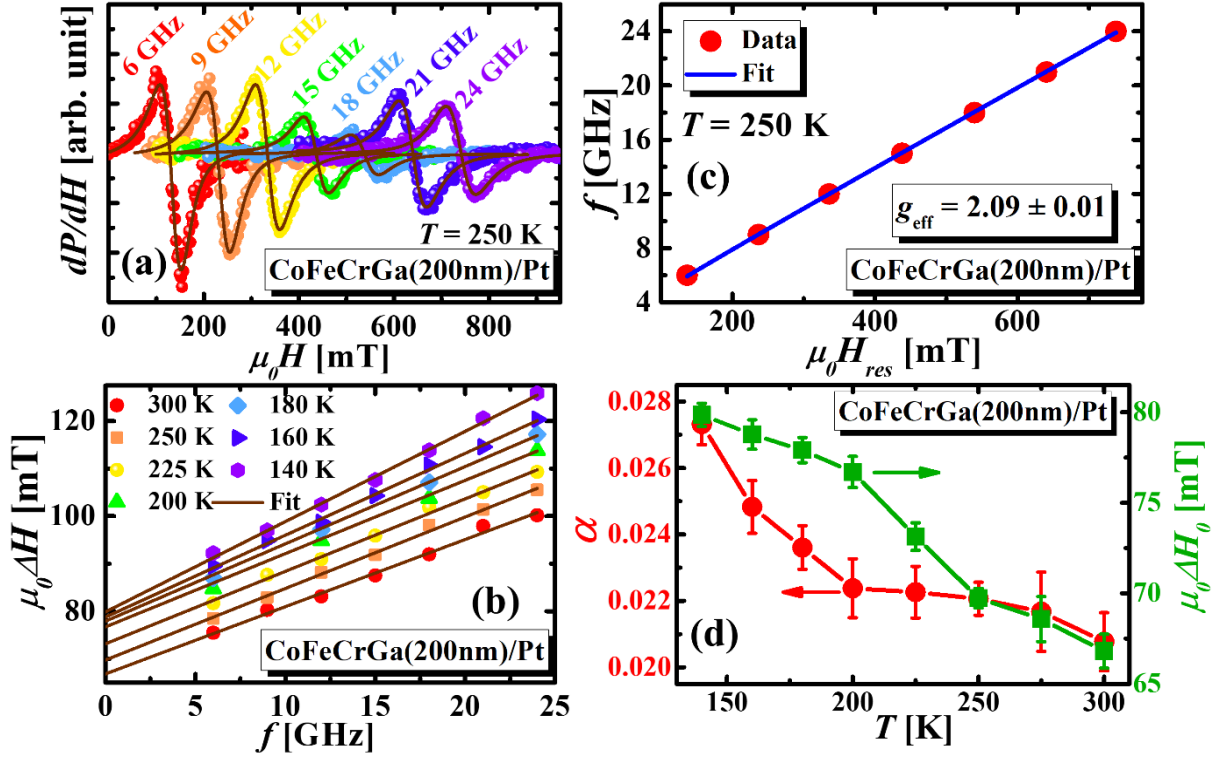

**Figure 10.** (a) Field-derivative of $\left(\frac{dP}{dH}\right)$ as a function of the IP DC magnetic field for various frequencies in the range: 6 GHz $\leq f \leq$ 24 GHz recorded at $T$ = 250 K for the MgO/CoFeCrGa(200nm)/Pt fitted with **Eqn. 7**. (b) The $\Delta H$ -$f$ curves for the MgO/CoFeCrGa(200nm)/Pt film at different temperatures fitted with $\Delta H = \Delta H_0 + \frac{4\pi\alpha}{\gamma\mu_0}f$. (c) Fitting of the $f$ vs. the resonance field, $H_{res}$ using the Kittel's equation at $T$ = 250 K for the MgO/CoFeCrGa(200nm)/Pt film. (d) Left $y$ scale: temperature dependence of damping parameter $\alpha(T)$ for the MgO/CoFeCrGa(200nm)/Pt , and right $y$-scale: temperature dependence of $\Delta H_0$ for the same.

The estimated value of $g_{eff} = (2.09 \pm 0.01)$ at 250 K for the MgO/CoFeCrGa(200nm)/Pt(5nm) film, which is slightly higher than the free electron value ($g_{eff}$ = 2.002). Note that $g_{eff} = (2.046 \pm 0.01)$ and $(2.048 \pm 0.02)$ for the MgO/CoFeCrGa(95nm) and MgO/CoFeCrGa(95nm)/Pt(5nm) films, respectively at 250 K. Finally, $\alpha(T)$ for the MgO/CoFeCrGa(200nm)/Pt film is shown on the left $y$-axis of **Fig. 10**(d). It is evident that $\alpha(T)$ increases with decreasing temperature, especially below 225 K similar to what we have observed for the MgO/CoFeCrGa(95nm) and MgO/

CoFeCrGa(95nm)/Pt(5nm) films. This observation further confirms the contribution of $\alpha$ towards the observed decrease in the LSSE signal in the CoFeCrGa films below the temperature range of 200-225 K.

## 4. CONCLUSIONS

In summary, we present a comprehensive investigation of the temperature ANE and intrinsic longitudinal spin Seebeck effect (LSSE) in the quaternary Heusler alloy based SGS thin films of CoFeCrGa grown on MgO substrates. We found that the anomalous Nernst coefficient for the MgO/CoFeCrGa (95 nm) film is $\approx 1.86\ \mu V.K^{-1}$ at room temperature which is much higher than the bulk polycrystalline sample of CoFeCrGa ($\approx 0.018\ \mu V.K^{-1}$ at 300 K) but comparable to that of the magnetic Weyl semimetal Co$_2$MnGa thin films ($\approx 2-3\ \mu V.K^{-1}$ at 300 K). Furthermore, the LSSE coefficient for our MgO/CoFeCrGa(95nm)/Pt(5nm) heterostructure is $\approx 20.5\ nV.K^{-1}.\Omega^{-1}$ at 295 K which is twice larger than that of the half-metallic ferromagnetic La$_{0.7}$Sr$_{0.3}$MnO$_3$ thin films ($\approx 9\ nV.K^{-1}.\Omega^{-1}$ at room temperature). We have shown that both ANE and LSSE coefficients follow identical temperature dependences and exhibit a maximum $\approx 225$ K which is understood as the combined effects of inelastic magnon scatterings and reduced magnon population at low temperatures. Our analyses not only indicated that the extrinsic skew scattering is the dominating mechanism for ANE in these films but also, provided critical insights into the functional form of the observed temperature dependent LSSE at low temperatures. Furthermore, by employing radio frequency transverse susceptibility and broadband ferromagnetic resonance in combination with the LSSE measurements, we have established a correlation among the observed LSSE signal, magnetic anisotropy and Gilbert damping of the CoFeCrGa thin films which will be beneficial for fabricating tunable and highly efficient spincaloritronic nanodevices. We believe that our findings will also attract the attention of materials science and spintronics community for further exploration of different Heusler alloys based magnetic thin films and heterostructures co-exhibiting multiple thermo-spin effects with promising efficiencies.

## ACKNOWLEDGEMENTS

HS and MHP acknowledge support from the US Department of Energy, Office of Basic Energy Sciences, Division of Materials Science and Engineering under Award No. DE-FG02-07ER46438. HS thanks the Alexander von Humboldt foundation for a research award and also acknowledges a visiting professorship at IIT Bombay. D.A.A. acknowledges the support of the National Science Foundation under Grant No. ECCS-1952957. DD and RC acknowledge the financial assistance received from DST Nanomission project (DST/NM/TUE/QM-11/2019).

## SUPPORTING INFORMATION

Magnetometry, temperature dependence of electrical resistivity, magnetic field and temperature dependences of transverse susceptibility, magnetic field dependence of ANE and LSSE voltages for the MgO/CoFeCrGa (200 nm) and MgO/CoFeCrGa (50 nm) films.

## DATA AVAILABILITY

The data that support the findings of this study are available from the corresponding author upon reasonable request.

# Supplementary Information

# Large thermo-spin effects in Heusler alloy based spin-gapless semiconductor thin films

Amit Chanda[1*], Deepika Rani[2], Derick DeTellem[1], Noha Alzahrani[1], Dario A. Arena[1], Sarath Witanachchi[1], Ratnamala Chatterjee[2], Manh-Huong Phan[1] and Hariharan Srikanth[1*]

[1] Department of Physics, University of South Florida, Tampa FL 33620

[2] Physics Department, Indian Institute of Technology Delhi, New Delhi - 110016

*Corresponding authors: achanda@usf.edu; sharihar@usf.edu

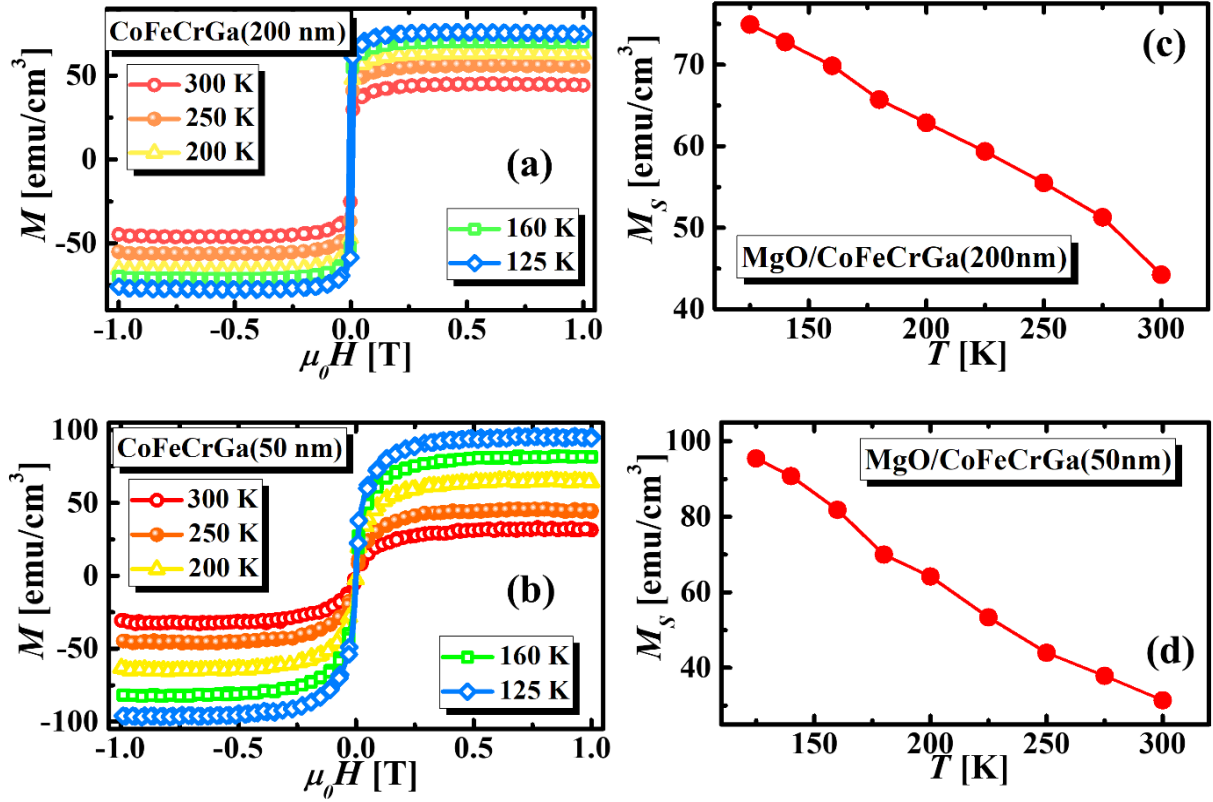

**Figure S1. (a) and (b)** Magnetic field dependence of magnetization, $M(H)$ of our MgO/CoFeCrGa(200nm) and MgO/CoFeCrGa(50nm)/Pt films. respectively measured at selected temperatures in the range: 125 K $\leq T \leq$ 300 K in presence of an in-plane sweeping magnetic field, (c) and (d) temperature dependence of the saturation magnetization, $M_S$ for the same films, respectively.

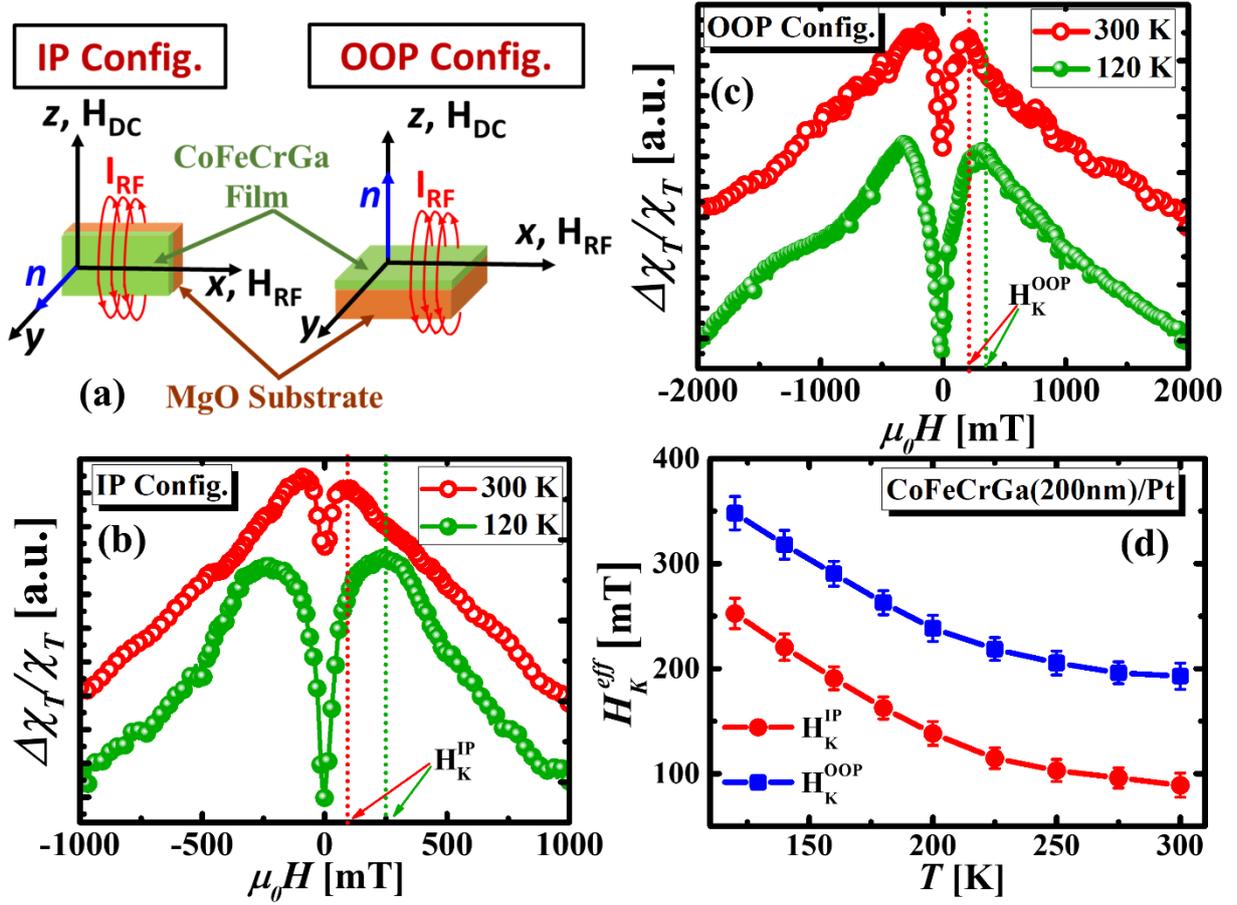

**Figure S2.** (a) Schematic illustration of the transverse susceptibilbity (TS) measurements. The bipolar field scans ($+H_{DC}^{max} \to -H_{DC}^{max} \to +H_{DC}^{max}$) of the field dependence of TS, $\frac{\Delta\chi_T}{\chi_T}(H_{DC})$ for the MgO/CoFeCrGa(200nm)/Pt(5nm) film measured at $T$ = 300 K and 20 K for the (b) IP ($H_{DC}$ is parallel to the film surface) and (c) OOP ($H_{DC}$ is perpendicular to the film surface) configurations. (d) Temperature variations of the effective anisotropy fields: $H_K^{IP}$ and $H_K^{OOP}$ for the MgO/CoFeCrGa(200nm)/Pt film.

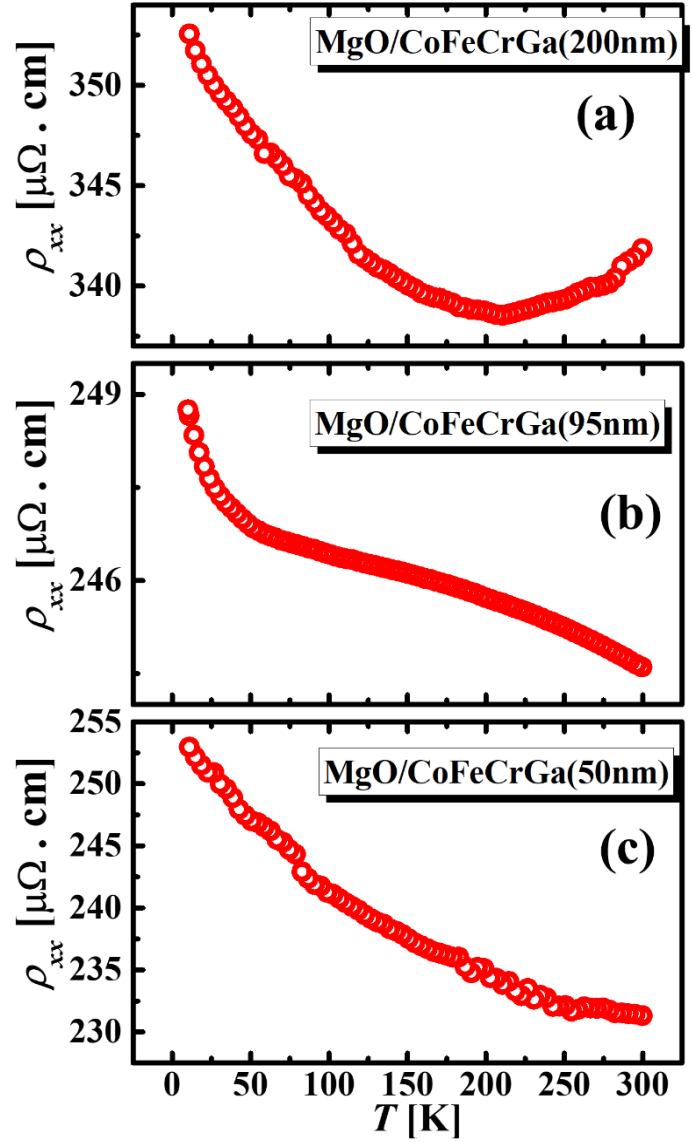

**Figure S3.** Temperature dependence of longitudinal resistivity, $\rho_{xx}(T)$ for the (a) MgO/CoFeCrGa(200nm), (b) MgO/CoFeCrGa(95nm) and (c) MgO/CoFeCrGa(200nm) films, respectively in the temperature range: 10 K $\leq T \leq$ 300 K.

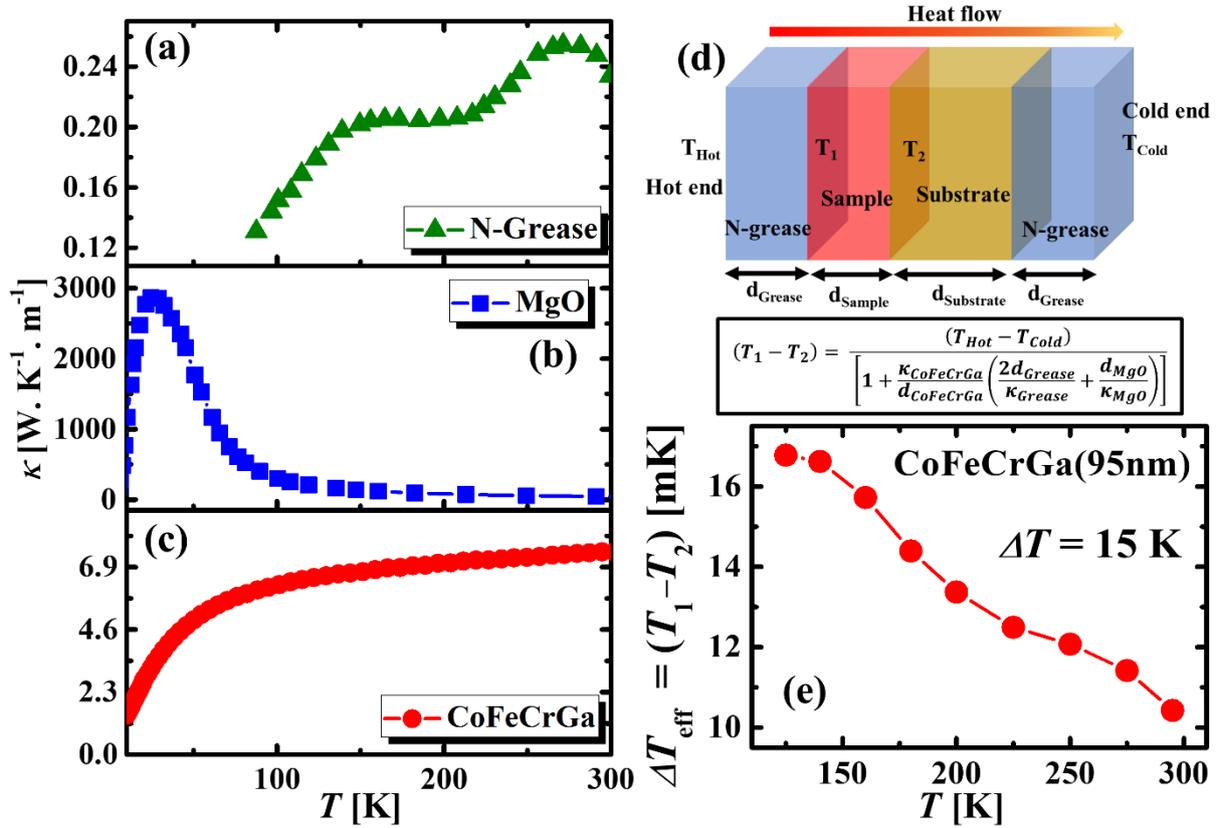

**Figure S4.** Temperature variations of thermal conductivity of (a) Apiezon N-grease,[1] (b) MgO crystal[2] and (c) bulk CoFeCrGa (measured) using the thermal transport option (TTO) of the PPMS. (d) Schematic illustration of the heat flow through N-grease/MgO substrate/CoFeCrGa film/N-grease considering the 4-slab model. (e) The temperture variation of the effective temperature difference across the MgO/CoFeCrGa(95nm) film estimated from the expression,[3]

$$\Delta T_{eff} = \Delta T_{\text{CoFeCrGa}} = \frac{\Delta T}{\left[1+\frac{\kappa_{CoFeCrGa}}{t_{CoFeCrGa}}\left(\frac{2t_{N-Grease}}{\kappa_{N-Grease}}+\frac{t_{MgO}}{\kappa_{MgO}}\right)\right]}$$

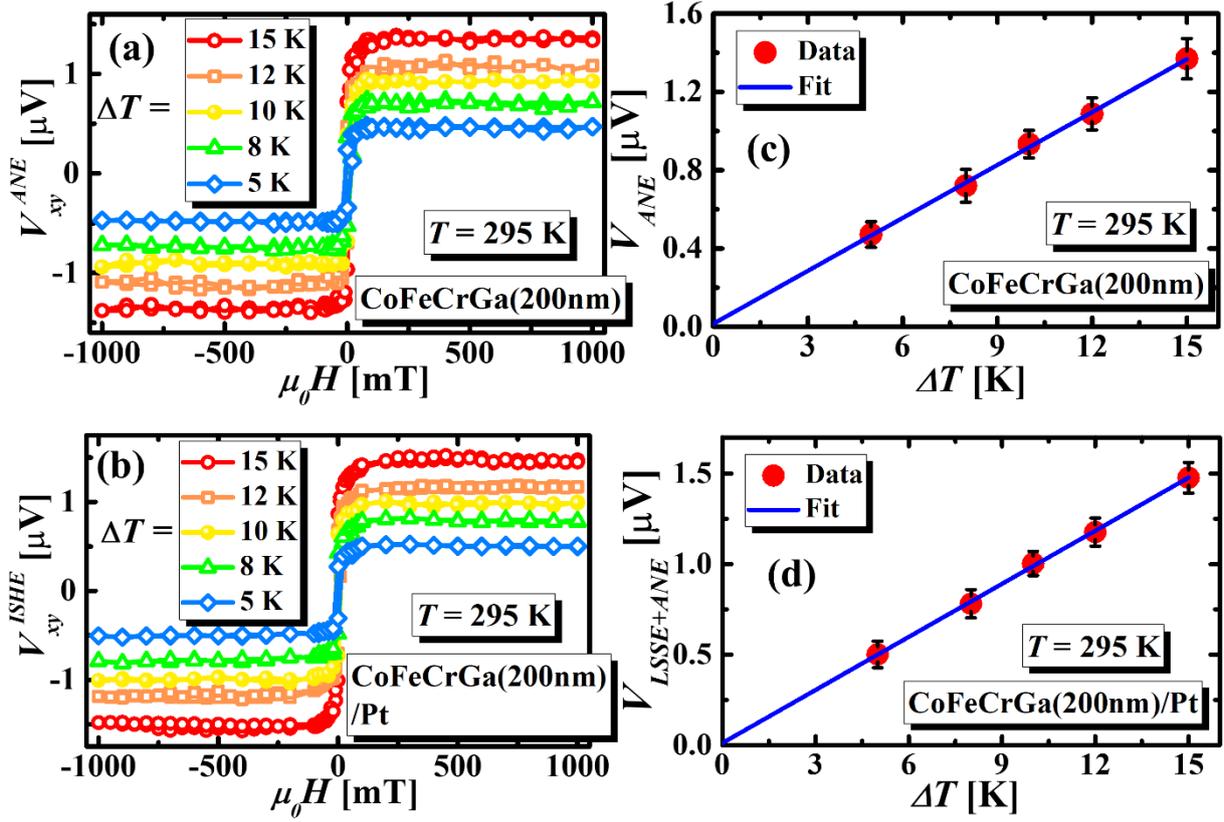

**Figure S5.** (a) and (b) The magnetic field dependence of the ANE voltage, $V_{ANE}(H)$ and ISHE-induced in-plane voltage, $V_{ISHE}(H)$ measured on the MgO/CoFeCrGa(200nm) and MgO/CoFeCrGa(200nm)/Pt films, respectively for different values of the temperature difference between the hot ($T_{hot}$) and cold ($T_{cold}$) copper blocks, $\Delta T = (T_{hot} - T_{cold})$ in the range: $+5\,\text{K} \leq \Delta T \leq +18\,\text{K}$ at a fixed average sample temperature $T = \frac{T_{hot}+T_{cold}}{2} = 295$ K. (c) and (d) The $\Delta T$ dependence of the background-corrected ANE voltage, $V_{ANE}(\Delta T) = \left[\frac{V_{ANE}(+\mu_0 H_{max},\,\Delta T)-V_{ANE}(-\mu_0 H_{max},\,\Delta T)}{2}\right]$ and the background-corrected (ANE+LSSE) voltage, $V_{ANE+LSSE}(\Delta T) = \left[\frac{V_{ISHE}(+\mu_0 H_{max},\,\Delta T)-V_{ISHE}(-\mu_0 H_{max},\,\Delta T)}{2}\right]$, respectively.

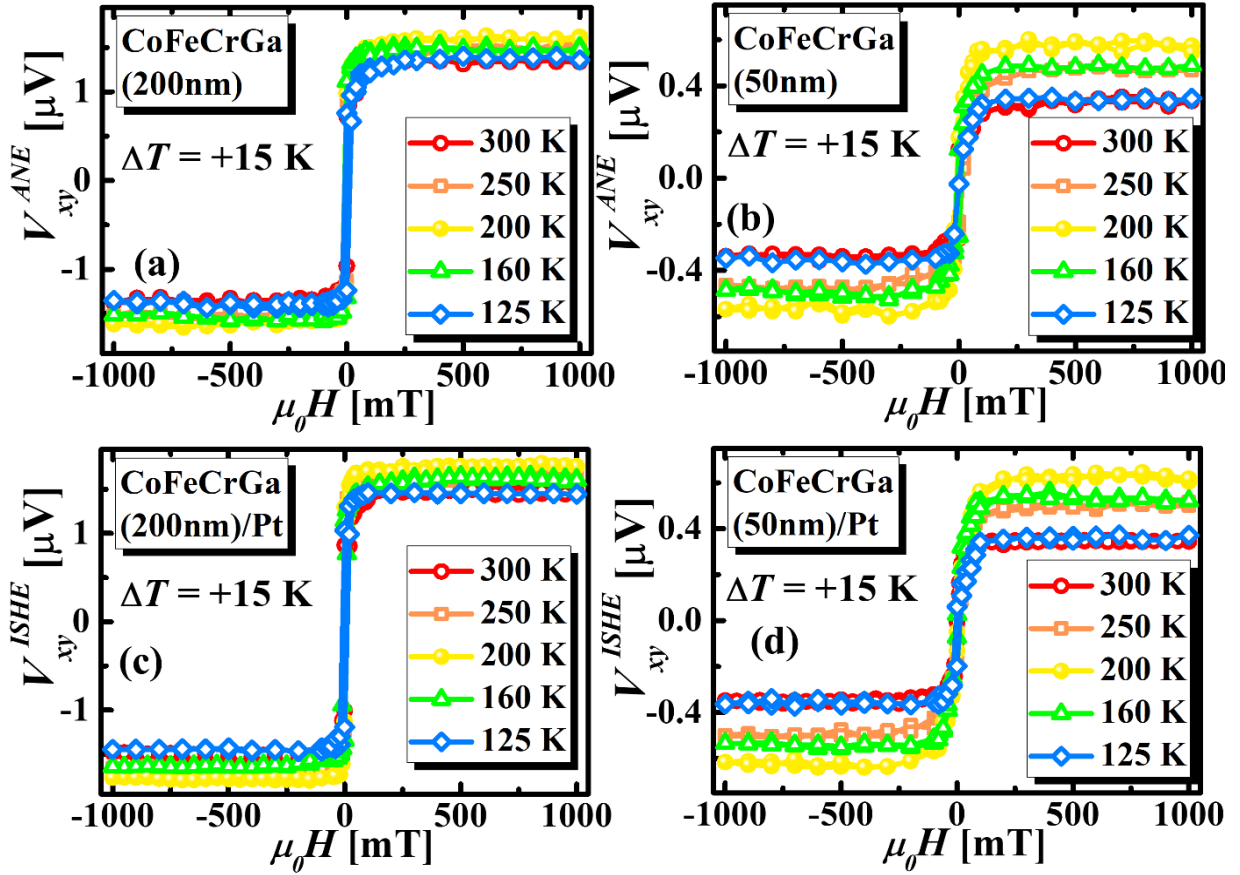

**Figure S6.** (a) and (b) $V_{ANE}(H)$ hysteresis loops measured at selected average sample temperatures in the temperature range: 125 K $\leq \Delta T \leq$ 295 K for a fixed value of $\Delta T$ = +15 K on the MgO/CoFeCrGa(200nm) and MgO/CoFeCrGa(500nm) films, respectively. (c) and (d) $V_{ISHE}(H)$ hysteresis loops measured at selected average sample temperatures in the temperature range: 125 K $\leq \Delta T \leq$ 295 K for a fixed value of $\Delta T$ = +15 K on the MgO/CoFeCrGa(200nm)/Pt and MgO/CoFeCrGa(50nm)/Pt films, respectively.